评述

# 6G 移动通信网络：愿景、挑战与关键技术

赵亚军 [1*]，郁光辉 [2]，徐汉青 [1]

1. 中兴通讯无线研究院算法部，北京，100029

2. 中兴通讯无线研究院算法部，深圳，518055

* 通信作者. E-mail: zhao.yajun1@zte.com.cn

**摘要** 随着 5G 网络开启规模商业部署，越来越多的研究机构及相关人员开始对下一代移动通信系统进行研究。本文将探讨十年后（2030 年～）的 6G 概念。本文首先用四个关键词概括未来 6G 愿景："智慧连接"、"深度连接"、"全息连接"和"泛在连接"，而这四个关键词共同构成"一念天地，万物随心"的 6G 总体愿景。接着分析了实现 6G 愿景面临的技术需求与挑战，包括峰值吞吐量、更高能效、随时随地的连接、全新理论与技术以及一些非技术挑战。然后分类给出了 6G 潜在关键技术：新频谱通信技术，包括太赫兹通信和可见光通信；基础性技术，包括稀疏理论（压缩感知）、全新信道编码、大规模天线及灵活频谱使用；专有技术特性，包括空天地海一体化网络和无线触觉网络。本文通过探讨 6G 愿景、需求与挑战以及潜在关键技术，尝试勾勒出 6G 的整体框架，以期为后续展开 6G 研究提供方向性指引。

**关键词** 6G, 愿景, 太赫兹, 可见光通信, 压缩感知, 全自由度双工, 无线触觉网络

## 1 引言

随着 5G 第一个标准版本完成，2019 年将会有 5G 网络设备小规模商用，首批符合 5G 标准的终端亦将上市。可以预期，拥有三大技术特性（enhanced Mobile BroadBand, eMBB; massive Machine-Type-Communications , mMTC ; ultra-Reliable Low-Latency Communications, uRLLC）的 5G 无线移动通信系统将支撑未来十年（2020～2030 年）信息社会的无线通信需求，成为有史以来最庞大复杂的通信网络，并将在多方面深刻影响社会发展及人类生活：与水和电一样，移动通信也将成为人类社会的基本需求；成为推动社会经济、文化和日常生活在内的社会结构变革的驱动力；将会极大地扩展人类的活动范围。

当然，上述 5G 愿景还需要通信领域的技术人员与其它相关行业人员一起努力，经过一定的时间逐步实现，包括标准不断完善、工程化逐步落地及商业应用模式突破等。这里从标准化角度观察 5G 标准不断成熟完善的过程。

当前处于 5G 标准化的第一阶段，即 5G 基础功能标准化阶段。此阶段主要针对 eMBB 技术特性优化，同时兼顾 uRLLC 和 mMTC 两种特性的基础需求，包括 5G NR Rel-15 和 Rel-16 两个标准版本。5G 第一个基础标准化版本（5G NR Rel-15）已基本完成，包括分阶段发布的三个子版本：国际标准化组织 3GPP 于 2018 年 3 月发布了第一个 5G 技术标准，支持非独立组网（Non-Standalone, NSA）与 eMBB 功能[1]；2018 年 9 月，3GPP 批准了 5G 独立组网 (Standalone, SA)技术标准[2]，5G 自此进入了产业全面冲刺的新阶段；2018 年 12 月 3GPP 于 RAN#82 全会上宣布，最后一个子版本（5G NR Rel-15 late drop）于 2019 年 3 月发布，支持 NR-NR DC（Dual Connectivity）特性[3]。而 5G 第二个标准版本（Rel-16），其所有技术特性已通过标准立项，相关标准化工作正在如火如荼的进行中，并将于 2019 年 12 月完成并正式发布。

5G 标准化的下一阶段（可称为"5G+"）将从 2020 年开始，对应的标准版本为 5G NR Rel-17 及后续版本，标准化重点包括两方面[4]：优化 uRLLC 和 mMTC 两种物联网（Internet of Things, IoT）特性，以更好支持垂直行业的应用（例如，工业无线互联网、高铁无线通信等）；设计支持 52.6GHz～114.25GHz 毫米波频段的空口特性。预期 5G 标准化的第二阶段将会吸引更多垂直行业领域成员参与标准制定，从而 5G 标准可以更好地针对垂直行业需求进行标准优化。

尽管 5G 尚处于规模商用起步阶段，相关技术特性还需要继续增强完善，物联网及垂直行业应用场景的业务模式也需要继续探索，但我们也有必要同步前瞻未来信息社会的通信需求，启动下一代移动通信系统概念与技术研究。这里我们尝试从三方面分析即刻启动下一代移动通信系统（为简化表达，下文将统一用 6G 标识）概念与技术研究的必然性。（1）十年周期法则。"自 1982 年引进第一个代（1G）移动通信系统以来，大约每十年更新一代无线移动通信系统"[5]，而且任何一代从开始概念研究到商业应用都需要十年左右的时间，也即，当上一代进入商用期，下一代开始概念和技术研究。5G 研究始于十年前，现在启动 6G 研究符合移动通信系统发展规律。（2）"鲶鱼效应"。不同于前几代移动通信系统，5G 主要针对物联网/垂直应用场景。随着 5G 网络规模部署，尤其是 5G 中后期，将会有众多垂直行业成员深度参与 5G 生态。与传统运营商主导的现状相比，未来新兴企业（尤其是天生具有创新思维的互联网公司）的深入参与将会对传统通信产业产生巨大冲击，甚至是革命性影响，即所谓"鲶鱼效应"。（3）IoT 业务模式爆发潜力。如同当年智能手机的出现刺激了 3G 应用并触发 4G 规模部署需求，相信 IoT 业务某些模式亦将会在 5G 时代某时间点刺激 5G 产业爆发，进而刺激对未来 6G 网络的需求。我们需要有足够的想象力，并需要为可能到来的未来网络提前着手准备，打好技术基础。综上分析，我们可以得出结论——现在是开启下一代无线移动通信系统（6G）研究的合适时机。

近期，越来越多的机构或个人开始涉及 B5G 或 6G 概念，包括学术界、工业界、政府甚至公众[6]-[9]。根据谷歌搜索引擎的统计，"6g technologies"是当今搜索量最大的 17 个关键词之一[10]。在 2018 美国移动世界大会上，美国联邦通讯委员会的一位官员在公开场合展望 6G[9]。不只美国，中国也已启动 6G 相关工作。2018 年 3 月工业和信息化部部长苗圩在接受媒体采访时表示，中国正着手研究 6G[11]。据悉，除中美两国外，欧盟、俄罗斯等也正在紧锣密鼓地开展相关工作。由此可以看出，业界对现在启动 6G 相关研究有一定的共识。

本文将主要探讨十年后（2030 年～）的通信需求和技术，即针对下一代无线移动通信系统（6G），其相对 5G 属于革命性（Revolution）系统。当然，不排除本文涉及的部分技术特性提前成熟或部分业务场景提前应用的可能，则本文把这部分归属 5G 演进特性（Evolution），即可以归属所谓 B5G（Beyond 5G）。可以预期，当前 5G 大部分特性将会在 6G 系统中继续保留并增强，但这些 5G 技术增强部分不属于本文讨论范畴。本文将侧重探

讨 6G 系统中可能引入的革命性关键技术。

本文将分别从需求驱动和技术驱动两个维度进行分析讨论, 重点探讨 6G 愿景、需求与挑战、潜在候选技术, 尝试勾勒出 6G 的整体框架, 以期为后续展开 6G 研究提供方向性指引。下文章节结构安排如下: 第 2 章, 畅想 6G 愿景与挑战; 第 3 章, 探讨可能用于 6G 的潜在关键技术; 最后在第 4 章给出全文总结。

## 2 6G 愿景与挑战

5G 启动初期, 确立的 5G 愿景为 "信息随心至, 万物触手及"[12]。基于此愿景, 确定了 5G 技术指标需求, 并进一步提出了候选关键技术。经历了概念确定、技术研究、标准化和产品开发过程, 5G 即将投入规模商用, 5G 愿景也将随着标准的完善及产业的成熟而逐步实现。现在要开启 6G 前瞻性研究, 也有必要首先确立 6G 愿景及相应的技术需求与挑战, 以牵引后续 6G 相关研究。5G 已经如此激动人心, 并将全面地赋能社会, 未来我们还能做什么?

本章将首先给出对未来 6G 愿景的畅想, 并浅析所述愿景的必然性, 然后进一步阐述实现 6G 愿景所面临的技术需求与挑战。

### 2.1 6G 愿景 (6G Vision)

当前 5G 的目标是渗透到社会的各个领域, 以用户为中心构建全方位的信息生态系统。但受限于标准化时间及相关技术发展的成熟度, 在信息交互的空间深度和广度上还有很多不足: 当前通信对象集中在陆地地表数千米高度的有限空间范围内; 虽然考虑了物联需求, 但距离真正无所不在的万物互联还有距离。尤其是随着人类活动范围的快速扩张, 众多技术领域的快速进步, 对更加广泛多样的信息交互提出了更高的需求。

6G 目标是满足十年后 (2030 年∼) 的信息社会需求, 因此 6G 愿景应该是现有 5G 不能满足而需要进一步提升的需求。基于 5G 可以满足的需求, 并结合其它相关领域的发展趋势, 我们认为 6G 愿景可以概括为四个关键词 (见图 1): **"智慧连接"**、**"深度连接"**、**"全息连接"**、**"泛在连接"**, 而这四个关键词共同构成 **"一念天地, 万物随心"** 的 6G 总体愿景。

5G 愿景 "信息随心至, 万物触手及", 强调信息交互、万物可连接, 而且连接对象集中在陆地 10km 高度的有限空间范围内。5G 虽然在 Rel-16 版本开始研究并标准化非陆地通信网络 (non-terrestrial networks, NTN) 技术特性[13], 但 NTN 架构涉及的卫星通信网络与蜂窝网络标准及技术体系依然是彼此独立, 需要通过专门的网关设备连接交互, 其通信能力和效率很难满足十年后的 "泛在连接" 需求。为满足未来 "泛在连接" 需求, 6G 需要引入下文所述的空天地海一体化网络, 该网络将是一个有机整体, 也即需要统一的标准协议架构和技术体系, 真正实现空天地海一体化的 "泛在连接"。另外, 5G 海量连接特性 (mMTC) 强调连接数量, 而不要求实时性; 超可靠低时延特性 (uRLLC) 强调可靠性与实时性, 但对连接数量和吞吐量没有需求, 属于以降低频谱效率和连接数量为代价实现的。而 6G 的 "万物随心" 愿景则同时需要海量连接、可靠性、实时性和吞吐量需求, 这些对通信网络是全新的巨大挑战, 其对应的典型场景为下文所述的无线触觉网络。因此, 虽然 6G 愿景涵盖的基本概念中部分在 5G 已有涉及, 但 6G 愿景提出了更高的目标, 以满足未来全新的场景需求。

概括来说, 6G 总体愿景是基于 5G 愿景的进一步扩展: "一念天地" 中的 "一念" 一词强调实时性, 指无处不在的低时延、大带宽的连接, "念" 还体现了思维与思维通信的 "深度连接", "天地" 对应空天地海无处不在的 "泛在连接"; "万物随心" 所指的万物为智能对象, 能够 "随心" 所想而智能响应, 即 "智慧连接"; 呈现方式也将支持 "随心" 无处不在的沉浸式全息交互体验, 即 "全息连接"。

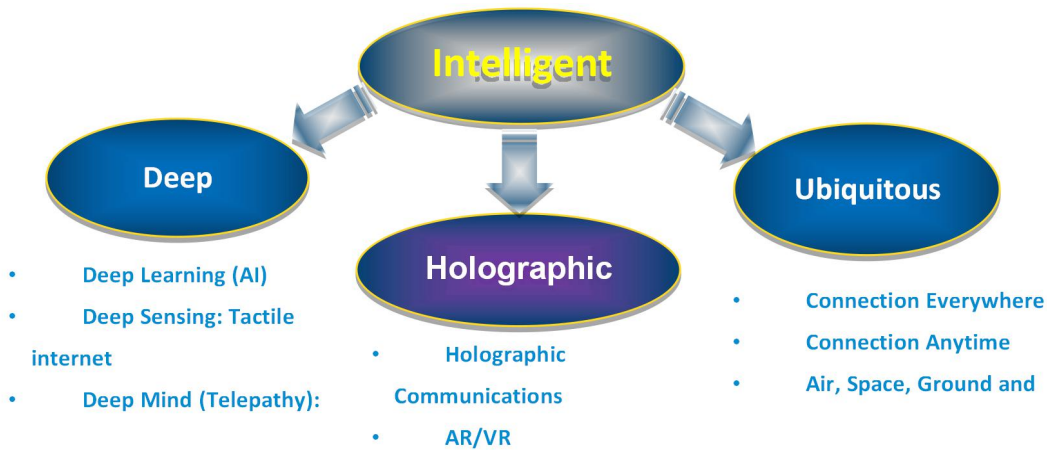

图 1 6G 愿景

Figure 1 6G Vision

● **智慧连接（Intelligent connectivity）**

人工智能（Artificial Intelligence，AI）是当前最热门的话题之一，几乎各个领域都在探索利用 AI 技术。无线移动通信网络与 AI 结合，让 AI 更好的赋能网络也成为必然趋势[14]-[30]。目前人们已经开始尝试在 5G 系统中使用 AI 技术[31]-[32]，但当前 5G 与 AI 的结合只能算是利用 AI 对传统网络架构进行优化改造，而不是真正以 AI 为基础的全新智能通信网络系统。首先，AI 技术应用于 5G 网络的时机相对较晚，最近几年才真正展开研究并尝试把 AI 技术应用在 5G 网络，而 5G 网络架构本身早已定型。尽管 5G 网络架构设计初期考虑了足够的灵活性（即所谓软件可定义），但毕竟没有考虑 AI 技术特点，依然算是传统的网络架构体系。其次，尽管 AI 技术发展很快，也已在一些领域展现了其强大的能力，但在更多领域依然处于探索阶段，AI 与无线通信技术结合研究更是刚起步不久，距离真正技术成熟还需要一个较长期的研究过程。

不过 AI 发展的趋势让我们看到了未来十年其技术成熟的可能性。同时，考虑到未来 6G 网络结构将会越来越庞大异构，业务类型和应用场景也越来越繁杂多变，充分利用 AI 技术来解决这种复杂的需求几乎是必然的选择。预期未来 6G 将会突破传统移动通信系统的应用范畴，演变成为支撑全社会、全领域/行业运行的基础性互联网络。若未来网络依然以现有统一的通信网络框架来支撑 6G 时代极度差异化的繁杂应用，将会面临前所未有的挑战。AI 技术的新一轮复兴及迅猛发展，为应对上述挑战并超越传统移动通信设计理念与性能提供了潜在的可能性，并将充分赋能未来 6G 网络[21]。因此，我们认为基于 AI 技术构建 6G 网络将是必然的选择，"智慧"将是 6G 网络的内在特征，即所谓"智慧连接"。

"智慧连接"特征可以表现为通信系统内在的全智能化：网元与网络架构的智能化、连接对象的智能化（终端设备智能化）、承载的信息支撑智能化业务。未来 6G 网络将会面临诸多挑战：更复杂、更庞大的网络，更多类型的终端及网络设备，更加复杂多样的业务类型。"智慧连接"将同时满足两方面的需求：一方面，所有相关连接在网络的设备本身智能化，相关业务也智能化；另一方面，复杂庞大的网络本身也需要智能化方式管理。"智慧连接"将是支撑 6G 网络其它三大特性"深度连接"、"全息连接"和"泛在连接"的基础特性。

● **深度连接（Deep connectivity）**

传统蜂窝网络（也包括即将规模部署的 5G 网络）已有深度覆盖的概念，主要是优化室内接入需求的深度覆盖。为实现室内深度覆盖，工程中一般采用室外宏基站覆盖室内，或者室内部署无线节点。4G 及之前几代的蜂窝网络系统是针对以人为中心的通信需求，深度覆盖针对人员活动的典型室内场景进行优化。经过多代无线通信系统的技术演进及工程经验积累，对人员活动场所的典型室内场景覆盖优化技术已经非常成熟。5G 开始，通信对象从以

人为中心的通信扩展为同时包括物联通信, 即所谓万物互联。因此, 5G 及未来无线通信网络设计及其部署需要同时兼顾人和物的深度覆盖需求, 尤其是物联场景的深度覆盖。

人类生产和生活空间不断扩大, 信息交互需求的类型和场景越来越复杂。以 5G 为开端的万物互联将会促进物联网通信需求快速提升, 并很可能在未来几年内爆发。相对人员的通信需求, 物联网信息交互无论是空间范围还是信息交互类型, 都将会极大的扩展。可以预期, 未来物联需求将会从几方面快速发展: (1) 连接对象活动空间的深度扩展。(2) 更深入的感知交互。未来的通信设备及其连接对象将大部分智能化, 从而需要更深度的感知、更实时的反馈与响应, 如同延伸的人类躯干和四肢; (3) 物理网络世界的深度数据挖掘。AI 深度学习将会对未来通信网络的数据深度挖掘与利用, 同时还包括为支持深度学习而强化的大数据通信需求; (4) 深入神经的交互。人机接口 (Brain Computer Interface, BCI) 等技术的成熟, 思维与思维的直接交互将成为可能, 一定程度的 "心灵感应" 将可能变为现实 [10][34]。因此, 我们预期十年后 (2030 年～) 的 6G 系统, 接入需求将从深度覆盖演变为 "深度连接 (Deep connectivity)", 其特征可以概括为如下几点:

- 深度感知 (Deep Sensing): 触觉网络 (Tactile Internet);
- 深度学习 (Deep Learning /AI): 深度数据挖掘;
- 深度思维 (Deep Mind): 心灵感应 (Telepathy)、思维与思维的直接交互 (Mind-to-Mind Communication)。

● **全息连接 (Holographic connectivity)**

AR/VR (Virtual and Augmented Reality) 被认为是 5G 最重要的需求之一, 尤其是对 5G 高吞吐量需求的典型应用之一, 5G 将能够支持把当前有线或固定无线接入的 AR/VR 变为更广泛场景的无线移动 AR/VR。一旦 AR/VR 可以更简单方便且不受位置限制的移动使用, 将会促进 AR/VR 业务快速发展, 进而刺激 AR/VR 技术与设备本身的快速发展与成熟。可以预期, 十年后 (2030 年～), 媒体交互形式将可能以现在平面多媒体为主, 发展为高保真 AR/VR 交互为主, 甚至全息信息交互, 进而无线全息通信将成为现实。高保真 AR/VR 将普遍存在, 全息通信及显示也可随时随地的进行, 从而人们可以在任何时间和地点享受完全沉浸式全息交互体验, 即实现所谓 "全息连接" 的通信愿景。当然, 若想基于无线通信网络实现全息通信、高保真 AR/VR 将会面临诸多挑战[35], 一系列文献已经在研究采用 AI 技术来解决相关问题[36]-[38], 即需要 "智慧连接" 的支持。

"全息连接 (Holographic connectivity)" 特征可以概括为: 全息通信、高保真 AR/VR、随时随地无缝覆盖的 AR/VR。

● **泛在连接 (Ubiquitous connectivity)**

传统蜂窝网络也有随时随地的无线接入需求。不过如前所述, 5G 系统开始, 相对人员的通信需求, 物联网信息交互无论是空间范围还是信息交互类型都将会极大的扩展。物联设备的活动范围将会极大扩展通信接入的地理空间, 包括布置于深地、深海或深空的无人探测器, 中高空有人/无人飞行器, 深入恶劣环境的自主机器人、远程遥控的智能机器设备等。另外, 随着宇航、深海探测等领域的科学技术快速发展, 在一些极端自然环境下的生存能力提升, 人类自身的活动空间也在快速扩展。例如, 2030～2040 年, 也许会有更多人有机会进入外太空, 则卫星与地面、卫星之间及与航天器之间的通信需求将会更普遍, 而不是现在仅仅局限于少数专业的科学探索领域的特殊通信需求; 人类在地面的活动踪迹也会更多的出现在极地、沙漠腹地等; 远洋的活动、更多无人岛屿进驻人类。上述通信场景构成十年后(2030年～) 更为广泛的 "随时随地" 连接需求, 即实现真正的 "泛在连接 (Ubiquitous connectivity)", "广阔" 的世界也将变得越来越触手可及。

"泛在连接" 特征可以概括为: 全地形、全空间立体覆盖连接, 即 "空-天-地-海" 随

时随地的连接，或称为空天地海一体化通信。对比"深度连接"和"泛在连接"，前者侧重连接对象的深度，后者强调地理区域的广度。

总结上述四大未来 6G 愿景，"智慧连接"是未来 6G 网络的大脑和神经，"深度连接"、"全息连接"和"泛在连接"三者构成 6G 网络的躯干，从而这四个特性共同使得未来 6G 网络成为完整的拥有"灵魂"的有机整体。未来通信系统将会在现有 5G 的基础上进一步发展增强，真正实现信息突破时空限制、网络拉近万物距离，实现无缝融合的人与万物智慧互联，并最终达到"一念天地，万物随心"的 6G 总体愿景。

## 2.2  需求与挑战（Requirements and Challenges）

上文 2.1 章节对未来 6G 网络做了畅想，其美好愿景让人无限期待。但若想实现这些美好的愿景，我们将不得不面临诸多技术需求与挑战。毫无疑问，5G 已有的几项基本技术指标还会在现有需求的基础上进一步提升，包括更高的吞吐量、更低的时延、更高的可靠性和更海量的连接数等。不过本文将重点讨论几项 6G 特有的关键技术需求与挑战。本节将会首先罗列这几项 6G 关键的技术需求与挑战，然后再对它们进行详细讨论和分析。

为实现 6G 网络的愿景，满足未来通信需求，如下几项关键技术需求与挑战需要被考虑（图 2）。

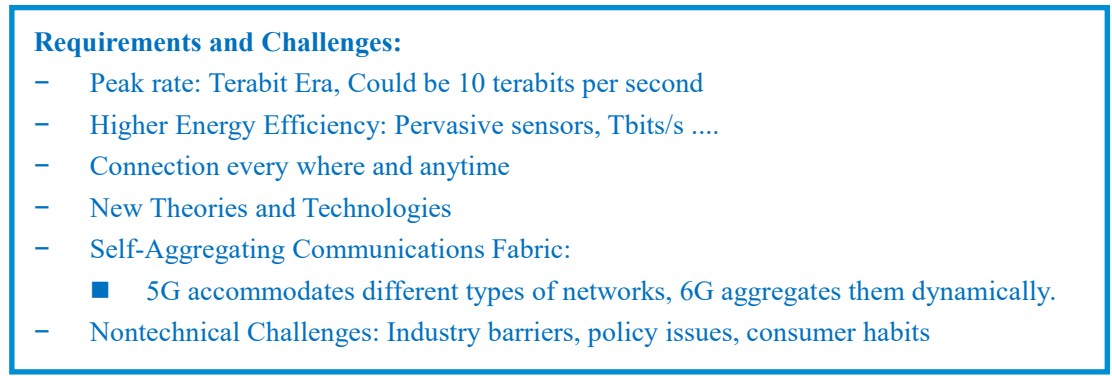

Requirements and Challenges:
- Peak rate: Terabit Era, Could be 10 terabits per second
- Higher Energy Efficiency: Pervasive sensors, Tbits/s ....
- Connection every where and anytime
- New Theories and Technologies
- Self-Aggregating Communications Fabric:
  - 5G accommodates different types of networks, 6G aggregates them dynamically.
- Nontechnical Challenges: Industry barriers, policy issues, consumer habits

图 2 6G 需求与挑战

**Figure 2** 6G Requirements and Challenges

● **峰值速率：太比特时代（Terabit Era，Tb/s）**

提及无线移动通信系统，人们首先要考虑的需求指标是峰值速率，峰值速率是从第一代无线移动通信系统开始就一直追求的关键技术指标之一。毫无疑问，6G 也必将进一步提升峰值速率。从无线通信系统发展规律和 6G 愿景两个角度分析可知，6G 峰值速率可能进入太比特时代（Terabit Era，Tb/s）。

首先，我们基于 1～5G 移动通信系统峰值速率提升的统计规律定量预测十年后（2030年～）的峰值速率需求。基于文献[44]的分析可知，1～5G 移动通信系统峰值速率的增长服从指数分布（按照各代系统标准化的时间点计算）。基于峰值速率对应文中表一第二列所示（1～5G 移动通信系统的峰值速率）预测未来十年的发展趋势，可知 2030 年可能达到 Tb/s 峰值速率。其次，从 6G 愿景定性分析可知，至少有两方面的应用需要 6G 峰值速率大幅度提升：（1）智能化（大数据）的普遍应用，需要海量的数据传输需求，基于大数据的智能化应用可能是触发下一代移动通信系统发展的重要驱动力之一；（2）高保真的 AR/VR 和全息通信将成为 6G 必然支持的应用，其所需的数据速率将远超我们目前已知的其他无线应用。

进一步，为达到高保真沉浸式 AR/VR，不仅需要 Tb/s 的峰值速率，还需要较低的交互时延，也即需要高吞吐率与低时延同时保证。另外，随时随地 AR/VR 意味着任何时间任何地点都希望可以满足高速率需求，也即不仅要求峰值速率，对网络平均速率和覆盖也有极高的

要求。

总结上述分析可知, 6G 网络将需要高达 Tb/s 级别的峰值速率。另外, 不同于以往仅要求局部覆盖区域 (例如热点区域) 的峰值速率需求, 6G 网络还将要求能够随时随地的享受高速率、低时延的连接需求, 这些将是 6G 网络需要面对的巨大挑战。

● **更高能效 (Higher Energy Efficiency)**

超大规模的移动通信网络已成为世界能源消耗的不可忽视的一部分。它不仅产生巨大的碳排放, 而且占据了相当一部分的运营成本。未来 6G 网络拥有超高吞吐量、超大带宽、超海量无处不在的无线节点, 这些将对能耗带来前所未有的巨大挑战。频谱效率提升和频谱带宽增大, 吞吐量可以有巨大的提升, 但随之而来的能效问题将会更加严重, 需要尽可能降低每比特的能量消耗 (J/bit)。无所不在、密集充满人类生产生活空间的感知网络传感器, 将带来两方面的能耗问题: 首先, 庞大的数量带来高昂的总能耗; 其次, 如何方便有效地对无处不在的部署进行供能也是挑战。另外, "智慧连接" 中海量数据处理功耗、超大规模天线的处理功耗等场景, 也是未来 6G 网络需要面临的功耗挑战。面对未来 6G 网络巨大的能源消费压力, 绿色节能通信显得尤为重要和迫切[45]。

● **随时随地的连接 (Connection Everywhere and Anytime)**

随着科学技术的进步, 人类活动空间将进一步扩大, 活动区域将更普遍的到达高空、外太空、远洋、深海; 通信节点, 尤其是物联节点相对人员将遍布更广阔的区域。通信网络已经和人类的社会活动密不可分, 未来需要构建一张无所不在 (覆盖空天地海)、无所不连 (万物互联)、无所不知 (借助各类传感器)、无所不用 (基于大数据和深度学习) 的网络, 真正实现随时随地的连接及交互需求。未来通信网络的通信目标应为: 任何人 (Anyone) 在任何时间 (Anytime) 任何地点 (Anywhere) 可与任何人 (Anyone) 进行任何业务 (Anyservice) 通信或与任何相关物体 (Related Objects) 进行相关信息 (Related Information) 交互[46]。

● **全新理论与技术 (New Theories and Technologies)**

为实现 6G 极具挑战性的愿景, 需要新增更多可用频谱资源, 同时也需要在一些基础性的理论与技术上有所突破。基于对 6G 愿景的需求分析, 我们认为需要在几个关键方面取得突破, 包括全新信号采样机制、全新信道编码与调制机制、太赫兹通信的理论与技术、AI 与无线通信结合的技术等。

● **自聚合通信架构 (Self-Aggregating Communications Fabric)**

几乎每一代 3GPP 标准都号称可以融合多种技术标准, 但最终结果依然还是一个自我封闭的标准系统。尽管 3GPP 标准希望包打天下, 但在万物互联逐渐实现的过程中, 我们将不得不面临与其它复杂多样的垂直行业标准和技术融合的问题。为更好支持万物互联及垂直行业应用, 6G 应该真正可以动态的融合多种技术体系, 具备对不同类型网络 (技术) 智能动态地自聚合能力。虽然 5G 能够一定程度地适应不同类型的网络 (技术), 但还是只能静态或半静态组合方式。6G 将需要实现以更加智能灵活的方式聚合不同类型的网络 (技术), 以动态自适应地满足复杂多样的场景及业务需求。

● **非技术性因素的挑战 (Nontechnical Challenges)**

未来 6G 若想顺利落地地实现, 不仅要面临上述技术性问题的挑战, 也将不得不需要尽力克服诸多非技术因素的挑战, 主要涉及行业壁垒、消费者习惯及政策法规问题等。

相对 5G, 6G 将会更加全面地渗透到社会生产、生活的各个方面, 与其它垂直行业领域的结合也将更加紧密。这意味着移动通信不再局限于自己的领域, 需要和其它垂直行业/领

域紧密配合。但是，一些传统行业固有的行为方式或利益关系将会对移动通信的进入直接或间接地设置行业壁垒。

频谱分配与使用规则是另一个非技术限制因素。例如 6G 太赫兹频段的使用，一方面需要全球不同国家和地区协调分配，尽可能分配统一的频段范围，同时还需要考虑与该频谱的其它领域使用者协调，例如气象雷达等。

卫星通信将面临更多的政策法规限制。首先，卫星通信所用的轨道资源、频谱资源等都需要各国协商解决。其次，相对传统地面通信，卫星通信在全球漫游切换方面上将面临更多挑战。目前，几个主要国家及一些商业实体都在积极进行卫星通信系统搭建，如何协调这些彼此独立部署的卫星通信系统关系，将是一个极其复杂的问题。

另外，移动通信进入大众多完全不同特点的垂直行业后，不得不面对差异化极大的用户使用习惯。如何更快速地改造这些千差万别的垂直行业用户固有思维方式和习惯，尽快适应全新的行为方式与规则，将是一个极具挑战的问题。

6G 网络最终将提供每秒太比特速率，支撑十年后（2030 年～）平均每人 1000+无线节点的连接，并提供随时随地的即时全息连接需求。未来将是一个完全的数据驱动的社会，人与万物被普遍地、近乎即时（毫秒级）地连接，构成一个不可思议的完全连接的乌托邦世界。

## 3 6G 候选关键技术

无线接入技术发展推动主要来自两个方面：关键理论/技术突破推动技术发展，应用需求驱动技术发展。对于未来 6G 会有哪些潜在的关键技术构成，不同的机构分别给出了不同的观点[6]-[10]。当前尚处于 6G 概念探讨的初期，各家给出的观点差异还比较大。但相信随着大家对 6G 概念探讨和技术研究的深入，认识将会逐渐清晰，研究方向也会不断收敛聚焦。本章将首先分类罗列 6G 潜在关键候选技术特性，然后对相关候选技术特性进行分析和解读。

为实现第 2 章所描绘的 6G 愿景及其挑战，同时考虑相关技术发展状况与趋势，我们认为 6G 潜在关键技术特性可以包括如下几方面（图 3）。

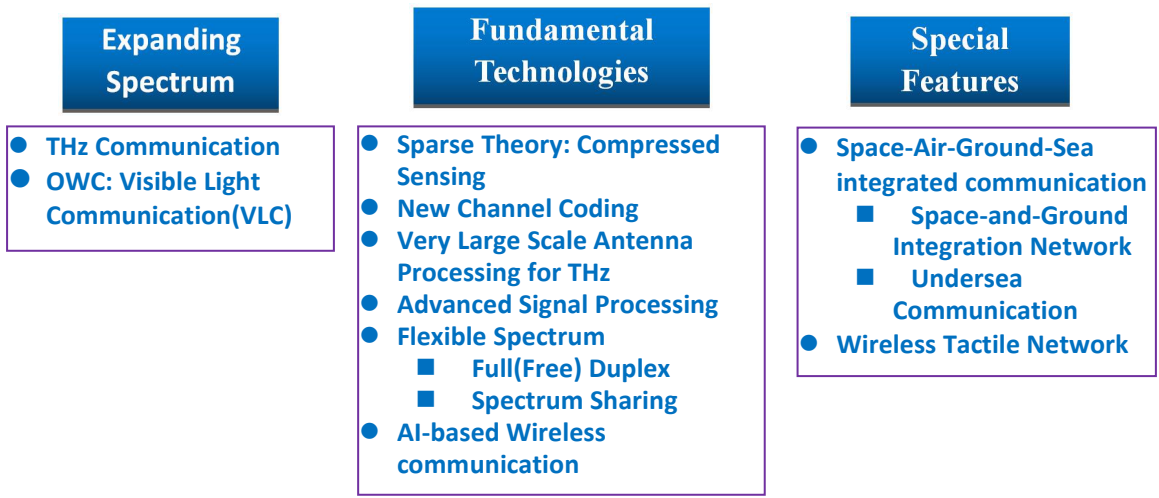

**图 3 6G 潜在关键技术特性**
**Figure 3** 6G Potential Key Technologies

基础性技术是构成 6G 网络的基石，只有关键基础技术被突破，6G 网络相应的技术需求才可能满足，进而相关愿景才可能实现。而专有技术特性则由多个关键的基础性技术点有机组成，用于满足未来 6G 典型场景的需求。从系统维度看，多个关键技术点组成专有技术特

性，而多个专有技术特性组合构建有机的系统。当前，我们需要对 6G 候选关键技术进行基础性研究和突破，为未来 6G 网络的标准化及工程实现技术研究奠定基础。其中，AI 与无线通信结合研究（"AI-based Wireless communication"）近期非常火热，也是实现未来 6G 网络"智慧连接"的关键技术，但是否可以作为无线领域的基础性技术尚存在争议。

### 3.1 新频谱通信技术

频谱是移动通信的基础，也是稀缺资源，持续增长的业务量需求要求未来移动通信系统扩展可用的频谱资源。太赫兹（Terahertz）和可见光（Visible Light）将是极具吸引力的两类重要的候选频谱。太赫兹频谱在通信等领域的开发利用受到了来自欧、美、日等国家和区域的高度重视，也获得了国际电信联盟（ITU）的大力支持。可见光通信技术是随着照明光源支持高速开关而发展起来的一种新型通信手段，可以有效的缓解当前射频通信频带紧张的问题，为短距离无线通信提供了一种新的选择方式。

本部分将分析太赫兹和可见光两类重要的候选频谱特性，探讨两者主要的应用场景，并给出面临的技术挑战。

#### 3.1.1 太赫兹通信（THz Communication）

太赫兹波是指频谱在 0.1～10 THz 之间的电磁波，波长为 30 至 3000 微米。频谱介于微波与远红外光之间，在其低波段与毫米波相衔，而在其高波段与红外光相衔，位于宏观电子学与微观光子学的过渡区域。太赫兹作为一个介于微波与光波之间的全新频段尚未被完全开发，太赫兹通信具有频谱资源丰富、传输速率高等优势，是未来移动通信中极具优势的宽带无线接入（Tb/s 级通信）技术[47]。美国联邦通信委员会专员 Jessica Rosenworcel 在 2018 年 9 月召开的美国移动通信世界大会上表示，6G 可以采用基于太赫兹（THz）频谱的网络和空间复用技术[9]。

太赫兹波以其独有的特性，使太赫兹通信比微波和无线光通信拥有许多优势，决定了太赫兹波在高速短距离宽带无线通信、宽带无线安全接入、空间通信等方面均有广阔的应用前景。（1）太赫兹波在空中传播时极易被空气中的水分吸收，比较适合于高速短距离无线通信；（2）波束更窄、方向性更好，具有更强的抗干扰能力，可实现 2～5 km 内的保密通信。（3）太赫兹波的频率高、带宽宽，能够满足无线宽带传输时对频谱带宽的需求。太赫兹波频谱在 108～1013GHz 之间，其中具有几十 GHz 的可用频谱带宽，可提供超过 Tb/s 的通信速率。（4）空间通信。在外层空间，太赫兹波在 350 μm、450 μm、620 μm、735 μm 和 870 μm 波长附近存在着相对透明的大气窗口，能够做到无损耗传输，极小的功率就可完成远距离通信。并且，相对无线光通信而言，波束更宽，接收端容易对准，量子噪声较低，天线终端可以小型化、平面化。因此，太赫兹波可广泛应用于空间通信中，特别适合用于卫星之间、星地之间的宽度通信。（5）太赫兹频段波长短，也适合采用更多天线阵子的 Massive MIMO（相对毫米波同样大小甚至更小的天线体积）。初步的研究表明，Massive MIMO 提供的波束赋型及空间复用增益可以很好的克服太赫兹传播的雨衰和大气衰落，可以满足密集城区覆盖需求（例如，200m 小区半径）。（6）能量效率高。相对于无线光通信而言，太赫兹波的光子能量低，大约是 10-3eV，只有可见光的 1/40，用它作为信息载体可以获得极高的能量效率。（7）穿透性强。太赫兹波能以较小的衰减穿透物质，适合一些特殊场景的通信需求。

太赫兹频段用于移动通信具有不可替代的优势，但同时面临着多方面的挑战：（1）覆盖与定位问题。电磁波传播特性表明，自由空间衰落大小与频率的平方成正比，因此太赫兹相对低频段有较大的自由空间衰落。太赫兹传播特性及巨量天线阵子，意味着太赫兹通信是高度定向的波束信号传播。我们需要针对这种高度定向传播的信号特征，重新设计和优化相关机制。（2）大尺度衰落特性。太赫兹信号对阴影非常敏感，对覆盖范围影响很大。例如，

如砖的信号衰减高达 40-80dB，人体可以带来 20-35dB 的信号衰减。不过湿度/降雨衰落对于太赫兹通信影响相对较小，因为湿度/降雨衰落在 100GHz 以下随着频率提升而快速增加，但在 100GHz 以上已经相对平坦。可以选择雨衰相对较小的几个太赫兹频段作为未来太赫兹通信的典型频段，例如 140GHz、220GHz 和 340GHz 等附近频段[47]。（3）快速信道波动与间歇性连接。给定的移动速度，信道相干时间与载波频率为线性关系，也即意味着太赫兹频段的相干时间很小，多普勒扩展较大，相比当前蜂窝系统所采用的频段变化快很多。此外，较高的阴影衰落将导致太赫兹传播的路径衰落更剧烈地波动。同时，太赫兹系统主要构成是小范围覆盖的微小区，而且是高度空间定向的信号传输，这意味着路径衰落、服务波束和小区关联关系将会迅速改变。从系统角度，意味着太赫兹通信系统的连接将表现为高度间歇性，需要有快速迅速适应机制来克服这种快速变化的间歇性连接问题。（4）处理功耗。利用超大规模天线的一个重大的挑战是宽带太赫兹系统模数（A/D）转换的功率消耗。功耗一般与采样率呈线性关系，而与每比特的采样数为指数关系。太赫兹频段大带宽和巨量天线需要高分辨率的量化，实现低功耗、低成本的设备将是巨大挑战。

为支持太赫兹通信，如下几方面需要进一步深入研究：（1）半导体技术，包括 RF、模拟基带和数字逻辑等；（2）研究低复杂度、低功耗的高速基带信号处理技术和集成电路设计方法，研制太赫兹高速通信基带平台；（3）调制解调，包括太赫兹直接调制、太赫兹混频调制和太赫兹光电调制等；（4）波形、信道编码；（5）同步机制，例如，高速高精度的捕获和跟踪机制、数百量级天线阵子的同步机制；（6）太赫兹空间和地面通信的信道测量与建模。上述几方面技术问题研究需要综合兼顾，以便在太赫兹通信的性能、复杂性和功耗之间取得平衡。

另外，在频谱监管方面，目前国际电联已决定将 0.12THz 和 0.2THz 划给无线通信使用，但 0.3THz 以上频段的监管规则尚不明晰，全球范围内尚未统一。需要国际电联层面和 WRC 会议共同努力，积极推动以达成共识。

太赫兹通信技术的研究只有二十年时间，很多关键器件还没有研制成功，一些关键技术不够成熟，还需进行大量的研究工作。但太赫兹通信是一个极具应用前景的技术，随着关键器件及关键技术的突破，太赫兹波通信技术必将给人类生产生活带来深远的影响。

### 3.1.2  可见光通信（Visible Light Communications）

一种对现有无线射频通信技术可能的补充技术是光无线通信（Optical Wireless Communications，OWC），频段包括红外、可见光和紫外，可以有效的缓解当前射频通信频带紧张的问题。其中，可见光频段是 OWC 最重要的频段，将在本节重点讨论。

可见光波段（390-700 纳米）的 OWC 系统通常被称为可见光通信（Visible Light Communications，VLC），它充分利用可见光发光二极管（LED）的优势，实现照明和高速数据通信的双重目的。与无线电通信相比，VLC 具有多方面极具吸引力的优势。首先，可见光通信技术可以提供大量潜在的可用频谱（THz 级带宽），并且频谱使用不受限，不需频谱监管机构的授权。其次，可见光通信不产生电磁辐射，也不易受外部电磁干扰影响，所以可广泛应用于对电磁干扰敏感、甚至必须消除电磁干扰的特殊场合，如医院、航空器、加油站和化工厂等。再次，可见光通信技术所搭建的网络安全性更高。该技术使用的传输媒介是可见光，不能穿透墙壁等遮挡物，传输限制在用户的视距范围以内，这就意味着网络信息的传输被局限在一个建筑物内，有效地避免了传输信息被外部恶意截获，保证了信息的安全性。最后，可见光通信技术支持快速搭建无线网络，方便灵活的组建临时网络与通信链路，降低网络使用与维护成本。像地铁、隧道等射频信号覆盖盲区，如果使用射频通信，则需要高昂的成本来建立基站，并支付昂贵的维护费用。而室内可见光通信技术可以利用其室内的照明光源作为基站，结合其它无线/有线通信技术，为用户提供便捷的室内无线通信服务。

OWC 典型应用场景包括：光热点（特别是在室内场景）、短距离通信、星间链路激光通信和海底通信（克服衰减和电磁干扰）。这些典型应用场景的 OWC 技术值得深入研究，并针对性的优化解决。

## 3.2 基础性技术

构成 6G 系统的潜在基础技术较多，本章将对其中最可能的潜在关键基础性技术展开讨论，包括稀疏理论（主要指压缩感知）、全新信道编码、大规模天线、灵活频谱技术等。

### 3.2.1 稀疏理论-压缩感知（Sparse Theory – Compressed Sensing）

信号采样是联系模拟信源和数字信息的桥梁。人们对信息的巨量需求对信号的采样、传输和存储带来巨大压力，如何缓解这种压力又能有效据取承载在信号中的有用信息是信号与信息处理中急需解决的关键问题之一。传统的信号处理是以香农-奈奎斯特（Shannon-Nyquist）采样定理为基础，信号通常先采样后压缩，而且必须以高于香农-奈奎斯特频率的速率对信号进行采样和处理。不同于香农-奈奎斯特信号采样机制，Donoho[48] 和 Candès、Tao、Romberg[49] 等人近年来基于信号稀疏性提出一种称为压缩感知/压缩采样（Compressed Sensing/Compressive Sampling, CS）的新颖采样理论，成功实现了信号的同时采样与压缩，为缓解上述压力提供了解决方法。CS 是获取、处理和恢复稀疏信号的有吸引力的范例[48]，这种全新模式是传统信息处理操作（包括采样、感知、压缩、估计和检测）极具竞争力的替代方案。此研究则挑战了香农-奈奎斯特采样定理[50]的理论极限，对整个信号处理领域产生了极其重要的影响。

CS 理论是当前信号处理领域的研究热点之一[48][49][51]–[57]。CS 的核心在于可以以计算有效的方式从欠定线性系统中恢复稀疏信号，即信号的少量线性测量（投影）包含用于其重建的足够信息。压缩感知理论指出：当信号在某个变换域是稀疏的或可压缩的，可以利用用与变换矩阵非相干的测量矩阵将变换系数线性投影作为低维观测向量，同时这种投影保持了重建信号所需的信息，通过进一步求解稀疏最优化问题就能够从低维观测向量精确地或高概率精确地重建原始高维信号。在该理论框架下，采样速率不再取决于信号的带宽，而是很大程度上取决于两个基本准则：稀疏性和非相干性，或者稀疏性和等距约束性。在压缩感知理论中，发端中信息采集（即数据观测或感知）代替了信号采样，而收端设备则用信号重建代替了传统的解码，因此不受香农-奈奎斯特采样率的限制。这一优势使得压缩感知在通信与信息处理的许多方面有着巨大的应用前景。

传统香农-奈奎斯特采样定理存在的问题：对于高宽带信号，香农-奈奎斯特采样定理需要至少两倍带宽的采样速率，对采样硬件设备要求较高；同时，产生的大量信号采样点对后续的传输及存储带来很重的负担，既浪费了大量的通信带宽资源又增加了通信设备成本；另外，也会因为要处理的数据量较多而降低信号处理的实时性。基于上文所述 CS 特性可知，利用 CS 特性完全可以克服传统香农-奈奎斯特采样定理的问题，更好的提升未来通信系统的性能：极大提升有用信息传输能力，降低有用信息传及处理时延。近年来，人们提出了利用目标信号稀疏性的各种无线通信应用，值得注意的例子包括信道估计、干扰抵消、方向估计、频谱感知和符号检测[52]。

压缩感知/稀疏理论在 5G 已有少量涉及，例如基于稀疏码的非正交多址（Sparse Code Multiple Access, SCMA）、Massive MIMO 的信道估计，但由于技术成熟度不足与标准化时间紧迫性的矛盾，最终并没有能够在 5G 标准中采纳。面对未来 6G 极具挑战的需求，压缩感知理论在 6G 中应用有更大的迫切性：下一代无线传输面临超大带宽、超大规模天线及超密集基站，将需要难以估量的计算复杂度、硬件成本及能量消耗；海量的物联网节点/触觉网络节点也需要利用压缩感知理论来解决信号采集压缩问题[51]。基于目前压缩感知/稀疏理

论研究的发展趋势，10 年后其技术成熟度完全可以满足工程化应用的需求，从而在 6G 系统中工程化落地成为可能。

结合 6G 将要面临的需求和挑战，有三种压缩感知典型应用场景：超宽带频谱感知、无线传感网络（无线触觉网络）、超大规模天线。

### 3.2.2 全新信道编码（New Channel Coding）

信道编码是无线通信的基础，下一代信道编码机制需要率先研究并突破，为未来 6G 无线通信系统打下基础。

相对目前 5G 系统，下一代信道编码机制研究需要满足未来更加复杂异构的无线通信场景和业务需求，需要考虑几方面的典型场景：超高吞吐量（Tb/s 级别）、超大带宽信道、超高频信道、可见光信道、高空/太空信道、远洋/深海信道、深地信道等复杂的传播环境及更异构多样的业务类型。

信道编码应用于未来无线通信系统同时涉及先进的信道编码算法和强大的芯片及实现技术两方面。前者受到后者工程实现的制约，因此需要对两者综合研究和突破。信道编码机制研究可以基于现有先进编码机制（如 Turbo、LDPC、Polar 等）获得适用于未来通信系统应用场景的基本信道编码原则，并进一步研究新的编解码机制及对应的芯片实现方案。需要对目前学术界正在研究的相关信道编码机制进行遴选，综合考虑其理论性能上限及对应的工程实现复杂度，选出下一代无线通信系统信道编码机制候选的突破方向。AI 在无线通信中的应用研究也给信道编码研究提供了一种全新的范式。经典的纠错码是根据编码理论设计的，而 AI 驱动的方法不再需要依赖于编码理论，为突破现有理论设计出全新信道编码机制提供了可能[58]。

另外，现有工程使用的信道编码设计假设为高斯点对点信道，而实际通信是多用户复杂网络场景的干扰/衰落信道，因此现有信道编码机制对实际干扰信道来说是次优的。未来通信网络干扰关系更加复杂，有必要考虑基于干扰信道假设进行优化设计，例如，多用户信道编码。

### 3.2.3 超大规模天线技术（Very Large Scale Antenna）

多天线技术，尤其超大规模天线技术，是提升无线移动通信系统频谱效率的关键技术之一。若想在未来 6G 网络中更好的发挥多天线的增益，我们将不得不面临诸多前所未有的需求和挑战。

从候选频谱角度，6G 极有可能采用太赫兹频谱通信。目前太赫兹频谱特性还未完全研究清楚，如何在太赫兹频谱上采用大规模天线更是面临诸多难题，包括工程理论突破和设计实现。同时，太赫兹频谱的引入也意味着未来通信系统频谱范围跨度更大，囊括 6GHz 以下低频、6GHz 以上毫米波及更高频太赫兹。另外，太赫兹频谱的大规模天线的阵子数量也会大幅增加，频谱效率要求更高。

面对 6G 需求的挑战，大规模天线技术需要研究并突破如下几方面的问题：解决跨频段、高效率、全空域覆盖天线射频领域的理论与技术实现问题；研究可配置、大规模阵列天线与射频技术，突破多频段、高集成射频电路面临的包括低功耗、高效率、低噪声、非线性、抗互扰等多项关键性挑战；提出新型大规模阵列天线设计理论与技术、高集成度射频电路优化设计理论与实现方法、以及高性能大规模模拟波束成型网络设计技术。

另外，为了充分得到大规模天线增益，需要在发射端和接收端获得信道状态信息（Channel Status Information, CSI）。即使假设 TDD 双工方式，依然会存在导频污染的问题，也即来自不同小区的上行链路导频序列彼此干扰。这些问题对于超大规模天线来说，即使仅仅为了获取不完美的 CSI 也是极具挑战性。尤其对于太赫兹频谱的大规模天线，其阵

子数更多, 需要估计的信道数目将会非常庞大。基于压缩感知理论对太赫兹频谱的大规模天线进行参考信号设计、信道估计与反馈是一个较好的选择, 包括 FDD 和 TDD 双工方式的 Massive MIMO 场景[59]-[61]。在 Massive MIMO 系统中[62][63], 发射机和/或接收机配备了大规模天线阵列, 由于散射群数量有限, 空间分辨率提高, 信道可以在角域中稀疏地表示[64][65][66]。另外, 相关研究和实际测量表明, 太赫兹信号到达由少量的路径簇构成, 且每个簇仅有较小的角度扩展。这些太赫兹频谱及其大规模天线的显著稀疏特性有利于采用压缩感知技术, 有效降低处理复杂度, 提升系统性能。

### 3.2.4 灵活频谱技术 (Flexible Spectrum)

上文所讨论的几项潜在的关键基础性技术都是为了进一步提升频谱效率, 使得频谱效率逼近信道容量上限, 从而在理想假设下达到网络峰值速率。而实际网络中, 更典型情况是频谱需求的不均衡性, 包括不同网络间的不均衡、同一网络内不同节点之间的不均衡、同一节点收发链路之间的不均衡等, 而这些不均衡特性导致频谱利用率低下。本章将分别探讨解决上述频谱需求不均衡问题的两种潜在候选技术: (1) 频谱共享, 主要用于解决不同网络间的频谱需求不均衡问题; (2) 全自由度双工, 主要用于解决同一网络内不同节点之间和同一节点收发链路之间的频谱需求不均衡问题。无线通信业务量需求激增与频谱资源紧缺的外在矛盾, 正驱动无线通信标准的内在变革。进一步提升频谱效率, 并消除对频谱资源利用方式的限制, 成为未来无线通信革新的一个目标。

#### 3.2.4.1 频谱共享 (Spectrum Sharing)

为满足未来 6G 系统频谱资源使用需求, 一方面, 需要扩展可用频谱, 例如采用太赫兹频谱和可见光频谱, 如 3.1 节所述; 另一方面也需要在频谱使用规则上有所改变, 突破目前授权载波使用方式为主的现状, 以更灵活的方式分配和使用频谱, 从而提升频谱资源利用率。目前蜂窝网络主要是采用授权载波的使用方式, 频谱资源所有者独占频谱使用权限, 即使所述频谱资源暂时空闲, 其它需求者也没有机会使用。独占授权频谱对用户的技术指标和使用区域等有严格的限制和要求, 能够有效避免系统间干扰并可以长期使用。然而, 这种方式在具备较高的稳定性和可靠性的同时, 也存在着因授权用户独占频段造成的频谱闲置、利用不充分等问题, 加剧了频谱供需矛盾。显然, 打破独占授权频谱的静态频谱划分使用规则, 采用频谱资源共享的方式是更好的选择[67]。

基于频谱资源授权方式划分, 频谱共享可以进一步分为两种类型: 非授权频谱, 用户使用频段不受限制, 彼此之间享有同等的使用权利但均不受到保护, 需要通过技术手段避免相互产生干扰; 动态共享频谱, 在保证主用户不受干扰的前提下, 通过设计许可权限 (如规定接入时间、接入地点、发射功率、干扰保护等), 赋予次用户相应的频谱使用权利, 次用户可使用数据库、频谱感知、认知无线电等技术, 在空间、时间、频率等不同维度上与主用户共享频谱。

对于非授权频谱, 目前主要的非授权载波频段包括 2.4GHz、5GHz, 占总可用频谱的比例较小, 不同国家和地区使用规则也不统一。WLAN 系统是最主要使用非授权载波的商业化的技术, 但频谱效率相对较低。3GPP LTE Rel-13 标准版本引入 LAA (Licensed-Assisted Access) 技术, 开创了蜂窝系统使用非授权载波的先例。当前, NR-unlicensed 技术特性正在 3GPP 5G 标准讨论中, 将会包含在 5G NR Rel-16 标准版本 (2019 年底完成并发布); 5G NR 也将可以利用非授权载波通信。而对于动态频谱共享, 尽管已有多年的研究, 但迄今尚未在规模商用网络中采用。

频谱共享技术没有被充分部署的原因有频谱分配规则约束的因素, 但更主要是频谱共享技术本身成熟度的限制。我们还是需要在频谱共享技术研究上有所突破, 包括高效频谱共享

技术及高效频谱监管技术，以在未来网络中更好的采用共享频谱技术提升频谱资源利用率，同时也可以更方便的进行频谱监管。频谱共享的实现技术可分为三大类：一是感知类，例如认知无线电技术（Cognitive Radio, CR）[68]；二是共享数据库类，如频谱池技术；三是将前两类技术结合起来使用。进一步，可以利用 AI 与频谱共享技术结合，以实现智能的动态频谱共享使用和智能的高效频谱监管[69]-[74]。

### 3.2.4.2  全自由度双工-全双工（Free Duplex - Full Duplex）

如上文所述，由于业务的数据包到达服从泊松分布，实际网络中收发链路（在蜂窝网络中一般指上下行链路）资源利用率动态波动，极不均衡。增强现有的双工技术是为了实现收发链路间灵活的频谱分配（或称为收发链路间灵活的频谱共享），从而从双工维度提升频谱资源利用率。

目前，相对传统的移动通信系统，5G 系统基于灵活空口概念设计，而双工方式则采用动态 TDD 架构，其中 FDD 模式仅仅是一种配置的特例。另外，5G 及后续 B5G/6G 主要可用频谱分布在 2GHz 以上的频段，这些频谱绝大部分为 TDD 频谱。解决上下行（Down Link / Up Link, DL/UL）交叉链路干扰的 CLI-RIM WID（Cross Link Interference - Remote Interference Management Work Item Description）标准项目将于 2019 年完成，并将包含在 5G NR Rel-16 标准版本[75]。此标准项目将会引入两类干扰抑制：解决相邻基站交叉链路干扰问题的机制，解决远端基站间交叉链路干扰（大气波导现象引起的交叉链路干扰）问题的机制。一旦这两类干扰被很好的解决，5G 将真正能够很好地支持灵活双工（Flexible Duplex）特性的商业部署，从而逐渐摆脱固定双工模式（Fixed Duplex, FDD/TDD）的资源利用限制。5G 初期的技术讨论虽涉及全双工技术，但由于其理论和技术研究尚不成熟，已没有在机会 5G 中采用。

随着未来十年双工技术的进步和工艺的成熟，预期 6G 时代的双工方式将有望实现真正**全自由度双工模式**（Free Duplex），即不再有 FDD/TDD 区分，而是根据收发链路间业务需求完全灵活自适应的调度为灵活双工或全双工（Full Duplex）模式，彻底打破双工机制对收发链路之间频谱资源利用的限制。全自由度双工模式通过收发链路（或 DL 与 UL）之间全自由度（时、频、空）灵活的频谱资源共享，将可以实现更加高效的频谱资源利用，达到提升吞吐量及降低传输时延的目的。而要实现全自由度双工模式，最关键的技术挑战是需要突破全双工技术。下图（图 5）描绘了无线移动通信系统双工方式的演进路线。

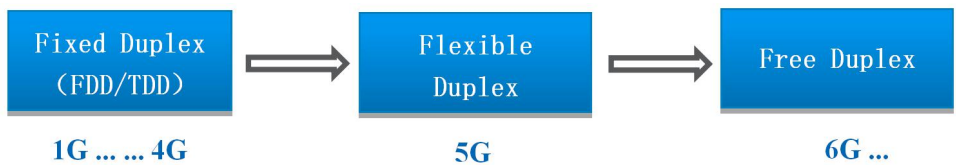

图 5 无线移动通信系统双工方式演进路线

**Figure 5** Duplex Evolution Route of Wireless Mobile Communication System

全双工可以最大限度的提升网络和接入设备收发设计的自由度，能够消除 FDD 和 TDD 资源使用限制，从而提升频谱效率和降低传输时延，可作为未来无线通信系统频谱提升的关键候选使能技术。

- 提升频谱效率：基于自干扰抑制技术的同时同频全双工可消除 FDD 和 TDD 资源使用限制，从理论极限上可提升一倍的频谱效率。
- 降低传输时延：未来的载波属性应该是以 TDD 载波为主。DL/UL 采用 TDD 方式传输，即使可以动态灵活上下行，甚至灵活时隙结构，依然存在上下行 TDD 带来的时延、切换操作等问题。全双工或者部分全双工，可以克服不能够同时传输带来的时延问

题, 同时对 DL/UL 资源调度提供更多的自由度、更大的灵活性。

同时同频全双工涉及的通信理论与工程技术研究已进行多年, 形成了空域、射频域、数字域联合的自干扰抑制技术路线。这些年很多研究机构已经成功设计出全双工收发机 [76][77], 并达到了 110dB 自干扰抑制能力[76]。全双工通信的应用领域十分广泛, 包括认知无线电系统[78]、中继网络[79][80]、双向通信系统[81]、终端与终端通信系统(Device to Device, D2D) [82]、蜂窝网[83][84][85]等。其中, 在蜂窝网尤其是覆盖范围小、发射功率低的密集蜂窝网场景的应用得到了越来越多的关注。

基于自干扰受限的技术特征可知, 全双工技术主要适合于如下几类典型应用场景: (1) 低发射功率场景, 包括短距离无线链路 (例如 D2D (Device to Device)、V2X (Vehicle to Everything)) 和小覆盖发射低功率的微小区 (Small Cell)。(2) 收发设备复杂度与成本不受限的场景, 例如无线中继 (Wireless Relay) 和无线回传 (Wireless Backhaul); (3) 窄波束且空间自由度较多的场景, 包括采用 Massive MIMO 的 6GHz 以下频段及高频毫米波/太赫兹频段的通信场景。

全双工技术的实用化进程中, 尚需解决的问题和技术挑战包括: 大功率动态自干扰信号的抑制、多天线射频域自干扰抑制电路的小型化、全双工体制下的网络新架构与干扰消除机制、与 FDD/TDD 半双工体制的共存和演进策略。另外, 从工程部署角度, 充分研究全双工的组网技术是更重要的方向。

### 3.2.4.2 基于 AI 的无线通信技术 (AI-based Wireless Communication)

近年来, 随着大数据时代的来临以及多种软硬件计算资源的增长, 人工智能 (AI) 特别是深度学习, 已经成为一个具有众多实际应用和活跃研究课题的领域。借助深度学习, 通过对数据进行深入归纳、分析, 从而获取新的、规律性的信息和知识, 并利用这些知识建立用于支持决策的模型, 进行风险分析或预测。深度学习的出现, 促进了许多领域快速发展, 例如语音认知、计算机视觉、机器翻译和生物信息等。而学术界和工业界也在不断思考如何将 AI 融入到无线通信系统中, 实现无线通信系统效能的大幅提升[86]-[87]。已有研究集中于应用层和网络层, 主要思想是将 AI 特别是深度学习的思想引入到无线资源管理与分配领域。不过, 该方向的研究正向 MAC 层和物理层推进, 特别在物理层出现无线传输与深度学习相结合的趋势。尽管无线大数据为 AI 与无线通信结合提供了可能, 但各项研究目前尚处于初步探索阶段, 智能通信系统的发展还需要一个长期的过程, 机遇与挑战共存[88]。

AI 在无线通信网络的应用层和网络层主要有两方面的应用。首先, 它们可以用于预测、推理和大数据分析。在此应用领域, AI 功能与无线网络从其用户、环境和网络设备生成的数据集学习的能力有关。例如, AI 可以用来分析和预测无线用户的可用性状态和内容请求, 从而使基站能够提前确定用户的关联内容并进行缓存, 从而减少数据流量负载。在这里, 与用户相关的行为模式 (如移动方式和内容请求) 将显著影响缓存哪些内容、网络中的哪个节点以及在什么时间缓存哪些内容。第二, AI 在无线网络中的另一个关键应用是通过在网络边缘及其各网元实体 (如基站和终端用户设备) 上内嵌 AI 功能来实现自组织网络操作。这种边缘智能是资源管理、用户关联和数据卸载的自组织解决方案的关键促成因素。在这种情况下, AI 可以学习环境, 并随着环境的变化采用不同的解决方案, 使得设备自主决策成为可能, 从而实现网络智能化[89]。当然, AI 可以同时用于无线通信网络的预测和自组织操作, 因为这两个功能在很大程度上是相互依赖的。

AI 用于物理层传输主要呈现出两种类型的深度学习网络, 一种基于数据驱动, 另一种基于数据模型双驱动。基于数据驱动的深度学习网络将无线通信系统的多个功能模块看作一个未知的黑盒子, 利用深度学习网络取而代之, 然后依赖大量训练数据完成输入到输出的训练。基于数据模型双驱动的深度学习网络在无线通信系统原有技术的基础上, 不改变无线通

信系统的模型结构，利用深度学习网络代替某个模块或者训练相关参数以提升某个模块的性能。AI 用于物理层传输，意味着底层基础的信号处理与通信机制等可能突破传统经典的通信理论框架，而采用基于 AI 驱动的信号处理及通信机制。不过，上述两种用于物理层传输的深度学习网络至少面临如下三方面的问题：

（1）基于深度学习的 AI 算法主要采用大量训练数据离线方式进行参数训练优化，而且由于训练数据获取的限制，一般为特定信道条件下的数据。这种处理机制产生了特定信道环境训练数据的离线静态训练与无线信道的多样性及动态时变性的矛盾；

（2）当前深度学习处理的为实数信号，而无线通信物理层传输的为复数信号。如何构建复数域的信号检测神经网络以更契合无线通信信号特点需要进一步研究；

（3）AI 用于物理层传输的训练样本主要采用数学仿真生成，仿真数据可能忽略了部分实际通信环境带来的影响。为了更好反应实际网络环境，需要利用更完备的、实际采集的数据进行相应网络的训练和测试。不过，如何有效获取足够的实际可信的训练数据是必须要解决的问题。例如，实际采样数据复杂多样，且存在大量虚警、错检数据，如何有效进行数据清理及合理分类将是巨大挑战。

为实现 6G 时代"智慧连接"的愿景，6G 网络将呈现为基于"分布式智能无线计算"（"Distributed intelligent wireless computing"[90]）网络架构以及基于 AI 的底层通信机制。也即，在 6G 时代，AI 将会被充分地集成到智能的 6G 网络系统中：

- AI 将在未来网络端到端的方方面面占据主导地位，包括：智能核心网和智能边缘网络，智能手机和智能物联网（超级物联网）终端，以及智能业务应用；
- 自主进化性能，如可用性、可修改性、有效性、安全性和效率；自主进化质量，如可测试性、可维护性、可重用性、可扩展性、可移植性和弹性；
- 底层基础的信号处理与通信机制将可能会突破传统经典的通信理论框架，全面采用 AI 驱动的机制。例如，基于深度学习的信道编译码[91]、基于深度学习的信号估计与检测[92]、基于深度学习的 MIMO 机制[93]-[95]、基于 AI 的资源调度[96]-[97]与分配[98]-[99]等。
- 网络基础设施具备自组织自优化能力，就像一个独立的自治系统。

随着 ICT 产业链架构融合的逐步深入、网络云化重构转型的加快以及更多新制式和技术的演进，电信运营商在网络运营方面将面临越来越大的压力和挑战，智能化网络是未来网络发展趋势，网络运营和运维模式将发生根本性变革。网络将由当前以人驱动为主的人治模式，逐步向网络自我驱动为主的自治模式转变。未来，智能化网络将通过网络数据、业务数据、用户数据等多维度感知，实现高度自治[100]。

## 3.3 专有技术特性

为实现上述 6G 网络的愿景与挑战，至少有两种潜在的关键专有技术特性需要被特别考虑，包括空天地海一体化通信和无线触觉网络。如前文所述，这些专有技术特性则由多个关键的基础性技术点有机组成，用于满足未来 6G 典型场景的需求，而这些专有技术特性组合构建为有机的 6G。本章将对这两种典型的专有技术特性进行较为详细的分析讨论。

### 3.3.1 空天地海一体化通信（Space-Air-Ground-Sea Integrated Communication）

空天地海一体化通信的目标是扩展通信覆盖广度和深度，也即在传统蜂窝网络的基础上分别与卫星通信（非陆地通信）和深海远洋通信（水下通信）深度融合。空天地海一体化网络是以地面网络为基础、以空间网络为延伸，覆盖太空、空中、陆地、海洋等自然空间，为天基（卫星通信网络）、空基（飞机、热气球、无人机等通信网络）、陆基（地面蜂窝网络）、海基（海洋水下无线通信+近海沿岸无线网络+远洋船只/悬浮岛屿等构成的网络）等各类用

户的活动提供信息保障的基础设施。从基本的构成上, 空天地海一体化通信系统可以包括两个子系统组成: 陆地移动通信网络与卫星通信网络结合的天地一体化子系统、陆地移动通信网络与深海远洋通信网络结合的深海远洋 (水下通信) 通信子系统。本章将分别探讨天地一体化通信和作为深海远洋通信最关键构成的水下无线通信。其中, 用于满足深海远洋通信场景的水下无线通信是否能够成为未来 6G 网络的组成部分存在争议, 本文仅是抛砖引玉, 尝试性提出来作为探讨。

### 3.3.1.1 天地一体化通信 (Space and Ground Integrated Communication)

天地一体化信息网络由卫星通信系统 (天基骨干网、天基接入网、地基节点网) 与地面互联网和移动通信网互联互通, 建成 "全球覆盖、随遇接入、按需服务、安全可信" 的天地一体化信息网络体系。下图提供了一个天地一体化网络架构参考例子 (图 6)。

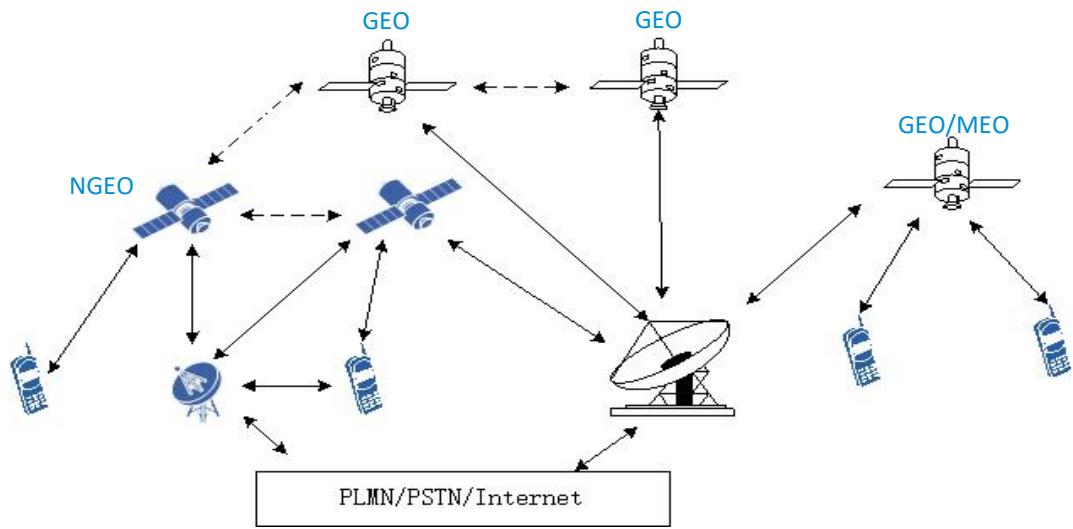

图 6 天地一体化网络架构

**Figure 6** Network architecture for space and ground integrated communication

文献 [101] 提供了一种典型的天地一体化网络架构, 可以作为未来 6G 网络天地一体化通信网络架构研究的参考。文中作者认为天基骨干网由布设在地球同步轨道的若干骨干卫星节点联网而成, 而骨干节点需要具备宽带接入、数据中继、路由交换、信息存储、处理融合等功能, 有单颗卫星或多个卫星簇构成; 天基接入网由布设在高轨或低轨的若干接入点组成, 满足陆海空天多层次海量用户的网络接入服务需求, 形成覆盖全球的接入网络; 同时, 地基节点网有多个地面互联的地基骨干节点组成, 主要完成网络控制、资源管理、协议转换、信息处理、融合共享等功能, 通过地面高速骨干网络完成组网, 并实现与其它地面系统的互联互通。

天地一体化网络特别是天基网络受到空间传播环境与网络设置等因素的影响, 与陆地移动通信网络存在显著差别 [102]: (1) 空间传输条件受限。空间节点由于距离遥远, 信道质量差, 链路通常存在较大的传输时延、较高的中断概率、非对称等特点; (2) 空间节点组网的特殊性。空间节点设置受轨道、星座等的制约, 节点高度动态、稀疏分布、拓扑结构动态变化等; (3) 系统组成与管理上的特殊性。有大量专用系统组成和专网构成, 各自长期发展中缺乏统一标准, 网络的管理实体应用需求和习惯也大相径庭, 不同管理域异构网络互联互通困难, 节点资源协同难。由于天基网络存在上述与陆地移动通信网络的显著差别, 大量陆地移动通信网络中的成熟技术难以直接用于天基网络。为尽快克服这些问题, 需要考虑从几方面入手: 尽快确定网络架构、确定接口标准、星间链路方案选择、天基信息处理、网络协议体系、安全机制等。

未来天地一体化通信网络五大典型的应用场景：（1）全地形覆盖：地面基站无法覆盖到的区域，如为海洋、湖泊、岛屿、山区等；移动平台，如飞机、远洋船舶、高铁。（2）应急通信：地震、海啸等灾害。（3）广播业务：低速的广播服务，如公共安全、应急响应消息等；广播，点播多媒体业务。（4）IoT 服务：远洋物资跟踪、偏远设备监控、大面积物联设备信息采集；（5）信令分流：通过卫星网络传递控制面的信息。

基于目前的发展状态，天地一体化网络还需要有如下几方面问题需要研究解决：传统卫星系统与移动通信网络的互联互通问题、卫星通信系统本身的技术突破问题、轨道与频谱资源分配管理问题、不同卫星系统之间的互联互通问题等。

### 3.3.1.2 水下无线通信（Undersea Communication）

水下无线通信是实现深海远洋通信的关键技术特性，可分为水下无线电磁波通信和水下非电磁波通信（主要包括水声通信和水下光通信）两种，它们分别具有不同的特性及应用场合。

● **水下无线电磁波通信（Undersea Wireless Electromagnetic Communication）**

电磁波是横波，在有电阻的导体中的穿透深度与其频率直接相关。频率越高，衰减越大，穿透深度越小。反之，频率越低，衰减相对越小，穿透深度越大。海水是良性的导体，趋肤效应较强，电磁波在海水中传输时会造成严重的影响，原本在陆地上传输良好的短波、中波、微波等无线电磁波在水下由于衰减的厉害，几乎无法传播。目前，各国发展的水下无线电磁波通信主要使用甚低频（Very Low Frequency，VLF）、超低频（Super Low Frequency，SLF）和极低频（Extremely Low Frequency，ELF）三个低频段。水下无线电磁波通信主要用于远距离的小深度的水下通信场景。

● **水声通信（Undersea Acoustic Communication）**

水声通信是其中最成熟的技术。声波是水中信息的主要载体，已广泛应用于水下通信、传感、探测、导航、定位等领域。声波属于机械波（纵波），在水下传输的信号衰减小（其衰减率为电磁波的千分之一），传输距离远，使用范围可从几百米延伸至几十公里，适用于温度稳定的深水通信。水声信道一个十分复杂的多径传输的信道，而且环境噪声高带宽窄可适用的载波频率低以及传输的时延大。为了克服这些不利因素，并尽可能地提高带宽利用效率，需要进一步研究新的技术方案。例如，多载波调制技术、多输入多输出技术。

● **水下无线光通信（Undersea Optical Wireless Communication）**

水下激光通信技术利用激光载波传输信息。由于波长 450nm～530nm 的蓝绿激光在水下的衰减较其他光波段小得多，因此蓝绿激光作为窗口波段应用于水下通信。蓝绿激光通信的优势是拥有几种方式中最高传输速率。在超近距离下，其速率可到达 100Mbps 级。蓝绿激光通信方向性好，接收天线较小。不过目前蓝绿激光应用于浅水近距离通信依然存在如下难点，需要进一步研究解决：（1）散射影响。水中悬浮颗粒及浮游生物会对光产生明显的散射作用，对于浑浊的浅水近距离传输，水下粒子造成的散射比空气中要强三个数量级，透过率明显降低。（2）光信号在水中的吸收效应严重。包括水媒质的吸收、溶解物的吸收及悬浮物的吸收等。（3）背景辐射的干扰。在接收信号的同时，来自水面外的强烈自然光，以及水下生物的辐射光也会对接收信噪比形成干扰。（4）高精度瞄准与实时跟踪困难。浅水区域活动繁多，移动的收发通信单元，在水下保持实时对准十分困难。并且由于激光只能进行视距通信，两个通信点间随机的遮挡都会影响通信性能。

### 3.3.2 无线触觉网络（Wireless Tactile Network）

当前 5G 网络所涉及的 IoT 网络主要是强调对万物的感知与连接, 而未来 6G 网络连接的对象将是普遍具备智能的对象, 其连接通信关系不仅是感知, 还包括实时的控制与响应, 即所谓 "触觉互联网"[103]。"触觉互联网" 指能够实时传送控制、触摸和感应/驱动信息的通信网络。IEEEP1918.1 标准工作组将触觉互联网定义为一个网络或一个网络的网络, 用于远程访问、感知、操作或控制感知实时的真实和虚拟对象或过程[104]。传统的互联网仅用于信息内容的交互, 而触觉互联网将不仅负责远程传递信息内容, 同时还包含与传递信息内容对应的远程控制与响应行为。它将提供从内容传递到远程技能集合传递的真正范式转换, 从而将可能革命性地改变社会的每个部分[104]–[108]。触觉互联网三个关键要素: 物理实时交互 (人和机器以感知的实时方式访问、操作和控制对象), 用于远程控制的超实时响应基础设施, 将控制和通信融入一个网络的应用程序上。

可以设想未来的物联世界: 大量的物联设备充满我们的环境, 执行各种传感任务; 这些设备可以随机部署, 以某种有组织的方式部署 (例如路边感应), 也可以作为普通智能手机平台的一部分; 然后这些设备都连接在一起, 通过各种复杂、相互不兼容的通信协议交换数据。这些物联设备/节点, 有些节点仅具有简单的感知功能, 有些节点则具有复杂的智能决策处理功能, 还有些节点负责动作响应。例如, 仅具有感知功能的物联设备构成所谓 "无线云" (Wireless Cloud), 在未来的无线网络中, 由非常廉价和低能耗的无线物联感知节点组成的密集集群协作, 为其他终端提供透明的通信服务。无线节点使用 "网络感知物理层" (network-aware physical layer) 进行操作, 该层处理它们接收到的叠加信号的混合, 并基于压缩感知的方式从中提取相关信息以转发到目的地。一些主要的应用场景可以包括: 远程机器人控制、远程机器操作、沉浸式虚拟现实、人际触觉通信、实时触觉广播、汽车和无人机控制等。可以预期, 6G 时代是无所不在的 "触觉互联网", 与无所不在的感知对象和/或智能对象进行实时传送控制、触摸和感应/驱动信息的通信, 从而实现 "一念天地, 万物随心"。

实现触觉互联网的关键技术挑战之一是将通信、控制和计算系统组合成一个共享的基础设施。通过将移动通信系统作为底层无线网络, 连同其软件化和虚拟化的逻辑网元实体, 集成为一个 (双向) 实时控制环路, 以使预期的实时控制与网络边缘高效计算能力相结合[109]。根据 ITU-T 关于触觉互联网的技术观察报告[110], 我们有必要进一步扩展无线触觉网络领域的研究, 包括引入全新的想法和概念, 提高接入网络内在的余度和分集, 以满足触觉互联网应用的严格延迟和可靠性要求。触觉互联网仍处于起步阶段。为了实现其愿景, 需要解决一些开放的研究挑战。除了波形选择和鲁棒调制方案等物理层问题外, 智能控制面与用户平面分离/协调技术对于减少信令开销和空中接口时延至关重要。为端到端降低时延, 有必要深入研究高度自适应网络编码技术和可扩展路由算法。另外, 对于触觉互联网来说, 安全性是最关键的需求之一, 必须要提供有效的保障机制以增强对恶意行为的防护。为确保规避未来可能的风险, 无线触觉网络的首要设计准则应该是通过授权辅助人类, 而不是自主地替代人类生产新的商品和提供服务[111]。

另外, 未来 6G 时代无所不在密集分布的传感器件将会产生海量的感知信息, 从而对无线网络容量带来巨大的挑战。另外, 传感器件进行海量信息采样处理能力、成本与能耗压力也是巨大挑战[112]。为克服这些海量信息处理需求的挑战, 采用压缩感知机制是很好的一个选择, 即无线触觉网络也是压缩感知机制使用的最典型场景之一[113]–[115]。无线传感器网络最直接的目标就是收集数据。由于传感器节点采集的数据有时空相关性, 满足压缩感知理论应用中信号是稀疏性和可压缩性的条件, 且传感器节点资源有限, 汇聚节点性能强大, 适用于压缩感知理论编码简单、解码复杂的特点, 因此, 基于压缩感知的无线传感网络 (Wireless Sensor Network, WSN) 数据收集的技术有了逐步深入和广泛的研究和发展。基于压缩感知机制的 WSN 有机会克服传统信号采集的问题, 有效实现无处不在的无线触觉网络

的应用需求。以传感网络定位信息为例，在无线传感器网络中，需要传感器节点的位置信息来执行位置感知、资源分配和调度。位置信息也是基于位置的服务的重要因素。由于位置网格中存在大量的元素，这种方法会产生很高的计算复杂度。由于目标节点的数目远小于网格中元素的数目，因此目标的位置信息是稀疏的，因此可以使用 CS 技术进行有效的定位[116][117][118]。

## 4 结论

本文用四个关键词概括未来 6G 愿景："智慧连接"、"深度连接"、"全息连接"和"泛在连接"，而这四个关键词共同构成"一念天地，万物随心"的 6G 总体愿景。分析了实现 6G 愿景所面临的技术需求与挑战，包括峰值吞吐量、更高能效、随时随地的连接、全新理论与技术以及一些非技术性因素的挑战。然后分类罗列并探讨了 6G 潜在关键技术：（1）新频谱通信技术，包括太赫兹通信和可见光通信；（2）基础性技术，包括稀疏理论（压缩感知）、全新信道编码、大规模天线及灵活频谱使用；（3）专有技术特性，包括空天地海一体化通信和无线触觉网络。

6G 愿景让人心潮澎湃，6G 关键候选技术充满挑战。6G 网络最终将提供每秒太比特速率，支撑十年后（2030 年～）平均每人 1000+无线节点的连接，并提供随时随地的即时全息连接需求。未来将是一个完全的数据驱动的社会，人与万物被普遍地、近乎即时（毫秒级）地连接，构成一个不可思议的完全连接的乌托邦世界。

**参考文献**

1　3GPP TSG RAN, "RP-180516 Report of 3GPP TSG RAN meeting #78", in Lisbon, Portugal, December 18 – 21, 2017.

2　3GPP TSG RAN, "RP-181999 Report of 3GPP TSG RAN meeting #80", in La Jolla, USA, June 11 – 14, 2018.

3　Balazs Bertenyi, Chairman 3GPP TSG RAN, "RAN adjusts schedule for 2nd wave of 5G specifications", http://www.3gpp.org/news-events/3gpp-news/2005-ran_r16_schedule , 3GPP RAN#82, December 14, 2018.

4　3GPP TSG RAN, "draft_MeetingReport_RAN_82_181213_eom", 3GPP RAN#82, December 14, 2018.

5　"Beyond　5G:　The　Roadmap　to　6G　and　beyond", https://www.cablefree.net/wireless-technology/4g-lte-beyond-5g-roadmap-6g-beyond/, Posted Jul 4, 2017Jan 18, 2018.

6　"Focus　Group　on　Technologies　for　Network　2030", https://www.itu.int/en/ITU-T/focusgroups/net2030/Pages/default.aspx

7　"University of Oulu Works On 6G", https://www.cdrinfo.com/d7/content/university-oulu-works-6g, Sep 17, 2018.

8　"东南大学教授尤肖虎: 6G 可能颠覆现有技术途径", http://news.seu.cn/2018/1217/c5485a255313/page.htm, 2018-12-15.

9　"FCC's Rosenworcel Talks Up 6G", https://www.multichannel.com/news/fccs-rosenworcel-talks-up-6g, SEP 14, 2018.

10　A. S. SWAROOP, GORI MOHAMED J, V. M. NIAZ AHAMED, FULL DUPLEX RADIO-WAVE TRANSMISSION FOR 6G INTERNET (6G connectivity). International Journal of Recent Trends in Engineering & Research (IJRTER), Volume 03, Issue 07; July – 2017 [ISSN: 2455-1457], DOI: 10.23883/IJRTER.2017.3349.GWZUR.

11　中国无线电管理, "苗圩部长: 我们已经在研究 6G 的发展", http://www.srrc.org.cn/article20461.aspx, 2018 年 03 月 12 日.

12　黄宇红, 王晓云, 刘光毅. 5G 移动通信系统概述 [J]. 电子技术应用, 2017(8).

13　3GPP TR 38.821 V0.4.0 (2019-03), "Solutions for NR to support non-terrestrial networks (NTN) (Release 16)".

14　N. C. Luong, D. T. Hoang, P. Wang, D. Niyato, D. I. Kim, and Z. Han, "Data collection and wireless communication in internet of things (IoT) using economic analysis and pricing models: a survey," IEEE Communications Surveys & Tutorials, vol. 18, no. 4, pp. 2546 - 2590, June 2016.

15　T. Park, N. Abuzainab, and W. Saad, "Learning how to communicate in the internet of things: Finite resources and heterogeneity," IEEE Access, vol. 4, pp. 7063 - 7073, Nov. 2016.

16　T. J. O' Shea and J. Hoydis, "An introduction to machine learning communications systems," available online arXiv:1702.00832, July 2017.

17　H. Sun, X. Chen, Q. Shi, M. Hong, X. Fu, and N. D. Sidiropoulos, "Learning to optimize: Training deep neural networks for wireless resource management," available online arXiv:1705.09412, May 2017.

18　C. Jiang, H. Zhang, Y. Ren, Z. Han, K. C. Chen, and L. Hanzo, "Machine learning paradigms for next-generation wireless networks," IEEE Wireless Communications, vol. 24, no. 2, pp. 98 - 105, April 2017.

19　M. Bkassiny, Y. Li, and S. K. Jayaweera, "A survey on machine-learning techniques in cognitive radios," IEEE Communications Surveys & Tutorials, vol. 15, no. 3, pp. 1136 - 1159, Oct. 2013.

20　N. Kato, Z. M. Fadlullah, B. Mao, F. Tang, O. Akashi, T. Inoue, and K. Mizutani, "The deep learning vision for heterogeneous network traffic control: Proposal, challenges, and future perspective," IEEE Wireless Communications, vol. 24, no. 3, pp. 146 - 153, Dec. 2017.

21  M. A. Alsheikh, S. Lin, D. Niyato, and H. P. Tan, "Machine learning in wireless sensor networks: Algorithms, strategies, and applications," IEEE Communications Surveys & Tutorials, vol. 16, no. 4, pp. 1996 - 2018, April 2014.

22  Y. Zhang, N. Meratnia, and P. Havinga, "Outlier detection techniques for wireless sensor networks: A survey," IEEE Communications Surveys & Tutorials, vol. 12, no. 2, pp. 159 - 170, April 2010.

23  M. Di and E. M. Joo, "A survey of machine learning in wireless sensor netoworks from networking and application perspectives," in Proc. of International Conference on Information, Communications & Signal Processing, Singapore, Dec. 2007.

24  R. M. Neal, Bayesian learning for neural networks, vol. 118, Springer Science & Business Media, 2012.

25  R. Rojas, Neural networks: a systematic introduction, Springer Science & Business Media, 2013.

26  H. B. Demuth, M. H. Beale, O. De Jess, and M. T. Hagan, Neural network design, Martin Hagan, 2014.

27  J. Schmidhuber, "Deep learning in neural networks: An overview," Neural networks, vol. 61, pp. 85 - 117, Jan. 2015.

28  J. G. Carbonell, R. S. Michalski, and T. M. Mitchell, "An overview of machine learning," in Machine learning, pp. 3 - 23.1983.

29  R. E. Schapire, "The boosting approach to machine learning: An overview," in Nonlinear Estimation and Classification, pp. 149 - 171. Springer, 2003.

30  E. Basegmez, "The next generation neural networks: Deep learning and spiking neural networks," in Advanced Seminar in Technical University of Munich, 2014, pp. 1 - 40.

31  You X H, Zhang C, Tan X S, et al. AI for 5G: research directions and paradigms (in Chinese). Sci Sin Inform, 2018, 48: 1589~1602, doi: 10.1360/N112018-00174

32  Kibria M G , Nguyen K , Villardi G P , et al. Big Data Analytics, Machine Learning and Artificial Intelligence in Next-Generation Wireless Networks[J]. IEEE Access, 2018, PP(99).

33  S. Bi, R. Zhang, Z. Ding, and S. Cui, "Wireless communications in the era of big data," IEEE Communications Magazine, vol. 53, no. 10, pp. 190–199, Oct. 2015.

34  "5G Is So Near-Future: A Look Ahead to 6G and 7G", https://iconsofinfrastructure.com/5g-is-so-near-future-a-look-ahead-to-6g-and-7g/ , Posted By Sylvie Barak, On February 16, 2018.

35  E. Ba̧stu̧g, M. Bennis, M. Médard, and M. Debbah, "Towards interconnected virtual reality: Opportunities, challenges and enablers," IEEE Communications Magazine, vol. 55, no. 6, pp. 110 - 117, June 2017.

36  J. Weissmann and R. Salomon, "Gesture recognition for virtual reality applications using data gloves and neural networks," in Proc. of International Joint Conference on Neural Networks (IJCNN), Washington, DC, USA, July 1999.

37  Deyou Xu, "A neural network approach for hand gesture recognition in virtual reality driving training system of SPG," in International Conference on Pattern Recognition, (ICPR), Hong Kong, China, Aug. 2006.

38  R. Hambli, A. Chamekh, and H. B. H. Salah, "Real-time deformation of structure using finite element and neural networks in virtual reality applications," Finite Elements in Analysis and Design, vol. 42, no. 11, pp. 985 - 991, July 2006.

39  J. Xie, R. Girshick, and A. Farhadi, "Deep3d: Fully automatic 2D-to-3D video conversion with deep convolutional neural networks," in European Conference on Computer Vision, Amsterdam, The Netherlands, Oct. 2006.

40  F. Qian, L. Ji, B. Han, and V. Gopalakrishnan, "Optimizing 360 video delivery over cellular networks," in Proc. of ACM Workshop on All Things Cellular: Operations, Applications and Challenges, NY, USA, Oct. 2016.

41  G. A. Koulieris, G. Drettakis, D. Cunningham, and K. Mania, "Gaze prediction using machine learning for dynamic stereo manipulation in games," in Proc. of IEEE Virtual Reality (VR), Greenville, SC, USA, March 2016.

42  M. Chen, W. Saad, and C. Yin, "Virtual reality over wireless networks: Quality-of-service model and learning-based resource management," available online: arxiv.org/abs/1703.04209, Mar. 2017.

43  M. Chen, W. Saad, and C. Yin, "Echo state transfer learning for data correlation aware resource allocation in wireless virtual reality," in Proc. of Asilomar Conference on Signals, Systems and Computers, Pacific Grove, CA, USA, Oct. 2017.

44  Raghavan V, Li J. Evolution of Physical-Layer Communications Research in the Post-5G Era[J]. 2019.

45  Ma Z, Zhang Z Q, Ding Z G, et al. Key techniques for 5G wireless communications: network architecture, physical layer, and MAC layer perspectives. Sci China Inf Sci, 2015, 58: 041301(20), doi: 10.1007/s11432-015-5293-y.

46  闵士权. "再论我国天地一体化综合信息网络构想." 卫星通信学术年会 2016.

47  Wells, J. Faster than fiber: The future of multi-G/s wireless[J]. IEEE Microwave Magazine, 2009, 10(3):104-112.

48  D. Donoho, "Compressed sensing," IEEE Trans. Inf. Theory, vol. 52, no. 4, pp. 1289-1306, April 2006.

49  E. Candes, J. Romberg, and T. Tao, "Stable signal recovery from incomplete and inaccurate measurements," Comm. Pure Appl. Math., vol. 59, no. 8, pp. 1207-1223, Aug. 2006.

50  A. V. Oppenheim and R. W. Schafer, Discrete-time Signal Processing, Prentice Hall, 2010.

51  Xiao S, Li T, Yan Y, et al. Compressed sensing in wireless sensor networks under complex conditions of Internet of things[J]. Cluster Computing, 2018(4):1-11.

52  Jun Won Choi∗, Byonghyo Shim♭, Yacong Ding♯, Bhaskar Rao♯, Dong In Kim♮, "Compressed Sensing for Wireless Communications: Useful Tips and Tricks", arXiv:1511.08746v3 [cs.IT] 20 Dec 2016.

53  Y. C. Eldar and G. Kutyniok, Compressed Sensing: Theory and Applications, Cambridge Univ. Press, 2012.

54  E. J. Candes and M. B. Wakin, "An introduction to compressive sampling," IEEE Signal Processing Mag., vol. 25, pp.21-30, March 2008.

55  Z. Han, H. Li, and W. Yin, Compressive Sensing for Wireless Networks, Cambridge Univ. Press, 2013.

56  Safarpour M, Inanlou R, Charmi M, et al. ADC-Assisted Random Sampler Architecture for Efficient Sparse Signal Acquisition[J]. IEEE Transactions on Very Large Scale Integration Systems, 2018, PP(99):1-5.

57  Szabo D, Gulyas A, Fitzek F H P, et al. Towards the Tactile Internet: Decreasing Communication Latency with Network Coding and Software Defined Networking[C]// European Wireless; European Wireless Conference. 2015.

58  Huang L, Zhang H, Li R, et al. AI Coding: Learning to Construct Error Correction Codes[J]. 2019.

59  Zhang R, Zhao H, Zhang J. Distributed Compressed Sensing Aided Sparse Channel Estimation in FDD Massive MIMO System[J]. IEEE Access, 2018, PP(99):1-1.

60  Kenarsari S R, Naeiny M F. Mobility-Aware Reconstruction Algorithm for Correlated Nonzero Neighborhood Structured Downlink channel in Massive MIMO[J]. IEEE Access, 2018, PP(99):1-1.

61  Akbarpour-Kasgari A, Ardebilipour M. Pilot allocation approaches for channel estimation in MIMO relay networks[J]. IET Communications, 2018, 12(16):2030-2037.

62  X. Rao, and V. K. N. Lau, "Distributed compressive CSIT estimation and feedback for FDD multi-user massive MIMO systems," IEEE Trans, Signal Process., vol. 62, no. 12, pp. 3261-3271, Jun. 2014.

63  X. Rao and V. K. N. Lau, "Compressive sensing with prior support quality information and application to massive MIMO channel estimation with temporal correlation," IEEE Trans. Signal Process., vol. 63, no. 18, pp. 4914-4924, Sept. 2015.

64  A. M. Sayeed, "Deconstructing multiantenna fading channels," IEEE Trans. Signal Process., vol. 50, no. 10, pp. 2563–2579, Oct. 2002.

65  D. Tse and P. Viswanath, Fundamentals of Wireless Communication, Cambridge Univ. Press, 2005.

66  H. L. Van Trees, Detection, estimation, and modulation theory, optimum array processing John Wiley & Sons, 2004.

67  Klaus David and Hendrik Berndt, "6G Vision and Requirements: Is There Any Need for Beyond 5G?", IEEE vehicular technology magazine, p1556–6072/18, September 2018.

68  B. Wang and K. J. R. Liu, "Advances in cognitive radio networks: a survey," IEEE Journal of Selected Topics on Signal Process., vol. 5, no. 1, pp. 5–23, Nov. 2010.

69  Cohen K, Nedić A, Srikant R. Distributed learning algorithms for spectrum sharing in spatial random access networks[C]. International Symposium on Modeling & Optimization in Mobile, Ad Hoc, & Wireless Networks. 2015.

70  Elhachimi, Jamal & Guennoun, Zouhair. (2014). An Artificial Intelligence Approach for Cognitive Spectrum Management. Int. J. Computer Technology & Applications, Vol 5 (3), 1219–1225. 5(3).

71  Bhattarai S , Park J M J , Gao B , et al. An Overview of Dynamic Spectrum Sharing: Ongoing Initiatives, Challenges, and a Roadmap for Future Research[J]. IEEE Transactions on Cognitive Communications and Networking, 2017, 2(2):110–128.

72  Lu J , Li L , Chen G , et al. Machine Learning based Intelligent Cognitive Network using Fog Computing[J]. 2017.

73  D. Cohen and Y. C. Eldar, "Sub-Nyquist sampling for power spectrum sensing in cognitive radios: a unified approach," IEEE Trans. Signal Process., vol. 62, no. 15, pp. 3897–3910, August 2014.

74  S. Romero and G. Leus, "Wideband spectrum sensing from compressed measurements using spectral prior information," IEEE Trans. Signal Process., vol. 61, no. 24, pp. 6232–6246, Dec. 2013.

75  LG Electronics, "RP-182864 Revised WID on Cross Link Interference (CLI) handling and Remote Interference Management (RIM) for NR (revision of RP-181652)", In: 3GPP TSG RAN Meeting #82, Sorrento, Italy, December 10–13, 2018.

76  Bharadia D, McMilin E, Katti S. Full duplex radios [J]. ACM Computer Communication Review, 2013, 43(4):375–386.

77  Duarte M, Dick C, Sabharwal A. Experiment-driven characterization of full duplex wireless systems [J]. IEEE Transaction Wireless Communications, 2012, 11(12):4296–4307.

78  LiaoYun, Song Lingyang,Han Zhu, etal. Full duplex cognitive radio: A new design paradigm for enhancing spectrum usage [J]. IEEE Communications Magazine,2015, 53(5): 138–145.

79  Liu Gang, Yu F, Ji Hong, et al. In-band full duplex relaying:A survey, researchissuesand challenges [J]. IEEE Communications Surveys Tutorials, 2015, 17(2): 500–524.

80  Riihonen T, Werner S, Wichman R. Hybrid full duplex/half duplex relaying with transmit power adaptation [J]. IEEE Transactions on Wireless Communications,2011, 10(9):3074–3085.

81  Ahmed E, Eltawil A, Sabharwal A. Rate gain region and design trade offs for full duplex wireless communications[J]. IEEE Transactions on Wireless Communications,2013,12(7):3556–3565.

82  Chai Xiaomeng, Liu Tong, Xing Chengwen, et al. Throughput improvement in cellular networks via full duplex based device to device communications [J]. IEEE Access,2016, 4(1):7645–7657.

83  Goyal S,Liu Pei,Panwar S, etal. Full duplex cellular systems: Will doubling interference prevent doubling capacity[J]. IEEE Communications Magazine,2015,53(5):121–127.

84  Nguyen D,Tran L,Pirinen P, etal. On the spectral efficiency of full duplex small cell wireless systems [J]. IEEE Transactions on Wireless Communications,2014,13(9):4896–4910.

85  Shao Shihai, Liu Donglin, Deng Kai, et al. Analysis of carrier utilization in full duplex cellular networks by dividing the co-channel interference region [J]. IEEE Communications Letters, 2014, 18(6): 1043–1046.

86  N. C. Luong, D. T. Hoang, P. Wang, D. Niyato, D. I. Kim, and Z. Han, "Data collection and wireless communication in internet of things (iot) using economic analysis and pricing models: a survey," IEEE Communications Surveys & Tutorials, vol. 18, no. 4, pp. 2546–2590, June 2016.

87  T. J. O'Shea and J. Hoydis, "An introduction to machine learning communications systems," available online arXiv:1702.00832, July 2017.

88  张静, 金石, 温朝凯, et al. 基于人工智能的无线传输技术最新研究进展[J]. 电信科学, 2018(8):46-55.

89  L. J. Lin, "Reinforcement learning for robots using neural networks," Tech. Rep., Carnegie-Mellon Univ Pittsburgh PA School of Computer Science, 1993.

90  Smartphones, "University of Oulu Works On 6G", https://www.cdrinfo.com/d7/content/university-oulu-works-6g, Sep 17, 2018.

91  Huang L , Zhang H , Li R , et al. AI Coding: Learning to Construct Error Correction Codes[J]. 2019.

92  Hao Ye, Geoffrey Ye Li, and Biing-Hwang Juang. Power of deep learning for channel estimation and signal detection in OFDM systems. IEEE Wireless Communications Letters, 7(1):114–117, 2018.

93  Timothy J O'Shea, Tugba Erpek, and T Charles Clancy. Deep learning based MIMO communications. arXiv preprint arXiv:1707.07980, 2017.

94  Mark Borgerding, Philip Schniter, and Sundeep Rangan. AMP-inspired deep networks for sparse linear inverse problems. IEEE Transactions on Signal Processing, 2017.

95  Takuya Fujihashi, Toshiaki Koike-Akino, Takashi Watanabe, and Philip V Orlik. Nonlinear equalization with deep learning for multipurpose visual MIMO communications. In Proc. IEEE International Conference on Communications (ICC), pages 1–6, 2018.

96  Qingchen Zhang, Man Lin, Laurence T Yang, Zhikui Chen, and Peng Li. Energy-efficient scheduling for real-time systems based on deep Qlearning model. IEEE Transactions on Sustainable Computing, 2017.

97  Ribal Atallah, Chadi Assi, and Maurice Khabbaz. Deep reinforcement learning-based scheduling for roadside communication networks. In Proc. 15th IEEE International Symposium on Modeling and Optimization in Mobile, Ad Hoc, and Wireless Networks (WiOpt), pages 1–8, 2017.

98  Haoran Sun, Xiangyi Chen, Qingjiang Shi, Mingyi Hong, Xiao Fu, and Nikos D Sidiropoulos. Learning to optimize: Training deep neural networks for wireless resource management. In Proc. 18th IEEE International Workshop on Signal Processing Advances in Wireless Communications (SPAWC), pages 1–6, 2017.

99  Zhiyuan Xu, Yanzhi Wang, Jian Tang, Jing Wang, and Mustafa Cenk Gursoy. A deep reinforcement learning based framework for powerefficient resource allocation in cloud RANs. In Proc. 2017 IEEE International Conference on Communications (ICC), pages 1–6.

100  Kibria M G , Nguyen K , Villardi G P , et al. Big Data Analytics, Machine Learning and Artificial Intelligence in Next-Generation Wireless Networks[J]. IEEE Access, 2018, PP(99).

101  李贺武, 吴茜, 徐恪, et al. 天地一体化网络研究进展与趋势[J]. 科技导报, 2016, 34(14):95-106.

102  张乃通, 赵康健, 刘功亮. 对建设我国"天地一体化信息网络"的思考[J]. 中国电子科学研究院学报, 2015, 10(3):223-230.

103  G. P. Fettweis, "The Tactile Internet: Applications and Challenges," IEEE Veh. Technol. Mag., vol. 9, no. 1, Mar. 2014, pp. 64 - 70.

104  ITU-T, " The tactile internet, " ITU-T Technology Watch Report [Online]. Available: https://www.itu.int/dms_pub/itut/oth/23/01/T23010000230001PDFE.pdf.

105  M. Dohler et al., "Internet of Skills, Where Robotics Meets AI, 5G and the Tactile Internet," EuCNC 2017.

106  G. Fettweis, "The Opportunities of the Tactile Internet—A Challenge for Future Electronics," http://www.lis.ei.tum.de/fileadmin/w00bdv/www/fpl2014/fettweis.pdf

107  M. Simsek, A. Aijaz, M. Dohler, J. Sachs, and G. Fettweis, "5G-Enabled Tactile Internet," invited submission to IEEE Journal on Selected Areas of Communication (JSAC), SI on Emerging Technologies, vol. 32, no. 3, pp. 1–14, 2016.

108  M. Dohler, T. Mahmoodi, M. Lema, and M. Condoluci, "Future of Mobile," EuCNC 2017.

109  IEEE 5G and Beyond technology roadmap - WHITE PAPER, https://5g.ieee.org/.

110  ITU-T Technology Watch Report, "The Tactile Internet," Aug. 2014.

111  E. Brynjolfsson and A. Mcafee, The Second Machine Age: Work, Progress,and Prosperity in a Time of Brilliant Technologies, W. W. Norton and Company, Jan. 2016.

112  Singh V K, Kumar M. A Compressed Sensing Approach to Resolve The Energy Hole Problem in Large Scale WSNs[J]. Wireless Personal Communications, 2018, 99(1):185–201.

113  Pagan J, Fallahzadeh R, Pedram M, et al. Toward Ultra-Low-Power Remote Health Monitoring: An Optimal and Adaptive Compressed Sensing Framework for Activity Recognition[J]. IEEE Transactions on Mobile Computing, 2018, PP(99):1–1.

114  Zhang J, Xiang Q, Yin Y, et al. Adaptive compressed sensing for wireless image sensor networks[J]. Multimedia Tools & Applications, 2017, 76(3):4227–4242.

115  Senel K , Larsson E G . Grant-Free Massive MTC-Enabled Massive MIMO: A Compressive Sensing Approach[J]. IEEE Transactions on Communications, 2018:1–1.

116  C. Feng, W. S. A. Au, S. Valalee, and Z. Tan, "Compressive sensing-based positioning using RSS of WLAN access points," IEEE Proc. INFOCOM 2010, pp. 1–9.

117  B. Zhang, X. Cheng, N. Zhang, Y. Cui, Y. Li, Q. Liang, "Sparse target counting and localization in sensor networks based on compressed sensing," IEEE Proc. INFOCOM 2011, pp. 2255–2263.

118  C. Feng, W. S. Au, S. Valalee, and Z. Tan, "Received-signal-strength-based indoor positioning using compressive sensing," IEEE Trans. Mobile Computing, vol. 11, no. 12, pp. 1983–1993, Dec. 2012.

# 6G Mobile Communication Network: Vision, Challenges and Key Technologies

Yajun ZHAO[1*], Guanghui YU[2], Hanqing XU[1]

1. *Algorithm Dept., Wireless Product R&D Institute, ZTE Corporation, Beijing* 100029*, China*；
2. *Algorithm Dept., Wireless Product R&D Institute, ZTE Corporation, Shenzhen* 518055*, China.*
* Corresponding author. E-mail: zhao.yajun1@zte.com.cn

**Abstract**  With the open of the scale-up commercial deployment of 5G network, more and more researchers and related organizations began to consider the next generation of mobile communication system. This article will explore the 6G concept for 2030s. Firstly, this article summarizes the future 6G vision with four keywords: "Intelligent Connectivity", "Deep Connectivity", "Holographic Connectivity" and "Ubiquitous Connectivity", and these four keywords together constitute the 6G overall vision of "Wherever you think, everything follows your heart". Then, the technical requirements and challenges to realize the 6G vision are analyzed, including peak throughput, higher energy efficiency, connection every where and anytime, new theories and technologies, self-aggregating communications fabric, and some non-technical challenges. Then the potential key technologies of 6G are classified and presented: communication technologies on new spectrum, including terahertz communication and visible light communication; fundamental technologies, including sparse theory (compressed sensing), new channel coding technology, large-scale antenna and flexible spectrum usage; special technical features, including Space-Air-Ground-Sea integrated communication and wireless tactile network. By exploring the 6G vision, requirements and challenges, as well as potential key technologies, this article attempts to outline the overall framework of 6G, and to provide directional guidance for the subsequent 6G research.

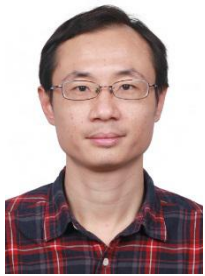

**Yajun ZHAO** was born in 1976. He has B.E. and Master degrees. From 2010 to the present time, he has acted as a radio expert in the wireless advanced research department in ZTE Corporation. Before that, he worked for Huawei on wireless technology research in wireless research department. At present, he is mainly engaged in the research of 5G standardization technology and future mobile communication technology (6G). His research interests include unlicensed spectrum, flexible duplex, CoMP and interference mitigation.

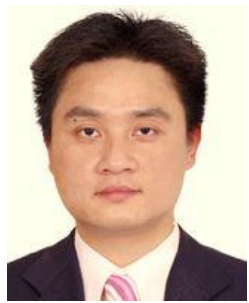

**Guanghui YU** received the Master and Ph.D. degrees in automatic control from Beijing Institute of Technology, Beijing China, in 1998 and 2003 respectively. Since 2003, he has acted as a radio expert in the wireless advanced research department in ZTE Corporation. His research interests include Wimax, 2G, 3G, 4G ,5G and B5G/6G design in RAN especially involved in multiplexing & access, MIMO, interference management, channel modeling as well as network architecture. He is also involved as one of the main researchers in the link and system simulation platform takes part in all kinds of 3GPP RAN1activities.

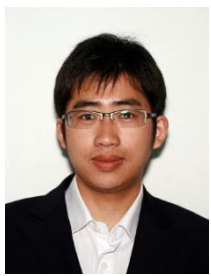

**Hanqing XU** was born in Yancheng city, China. He obtained his B.E. degree in communication engineering, and M.E. degree in communication and information system in 2006 and 2009. From 2009 to the present time, as an algorithm engineer of ZTE Corporation, his main research fields include small cell enhancement, flexible duplex and communication system operation at unlicensed spectrum in 3G/4G/5G system.